\newif\ifshortversion

\shortversiontrue 

\ifshortversion
	\documentclass[runningheads]{llncs}
	\renewcommand{\keywordname}{{\bfseries Palavras-chave:}}
\else
	\documentclass[a4paper,UKenglish,portuguese,cleveref,autoref]{lipics-v2019}
\fi

\usepackage[portuguese]{babel}
\usepackage{caption}
\usepackage{cite}
\usepackage{float}
\usepackage{graphicx}
\usepackage[utf8]{inputenc}
\usepackage{listings}
\usepackage{ocamllang}
\usepackage{textcomp}
\usepackage{url}

\makeatletter
\newcommand{\@chapapp}{\relax}%
\makeatother

\usepackage{caption}
\usepackage{subcaption}
\usepackage{url}

\usepackage[breaklinks]{hyperref}
\usepackage[breaklinks]{hyperref}

\begin{document}
\title{Visualização e animação de autómatos em Ocsigen Framework%
\thanks{Trabalho parcialmente suportado pela Fundação Tezos através do
projeto FACTOR (\url{http://www-ctp.di.fct.unl.pt/FACTOR/}) e por fundos nacionais através da FCT – Fundação para a Ciência e a Tecnologia, I.P., no âmbito do NOVA LINCS através do projeto UID/CEC/04516/2019}}
%
%
\author{Rita Macedo, Artur Miguel Dias, António Ravara}
\authorrunning{R. Macedo, A. M. Dias, A. Ravara}
%
\institute{NOVA LINCS e DI-FCT, Universidade NOVA de Lisboa}
\maketitle              
\begin{abstract}

Linguagens Formais e Teoria de Autómatos são bases importantes na formação em Engenharia Informática. O seu carácter rigoroso e formal torna exigente a sua aprendizagem. Um apoio importante à assimilação dos conceitos 
é a possibilidade de se visualizar interactivamente exemplos concretos destes modelos computacionais, facilitando a compreensão dos mesmos. As ferramentas disponíveis não são completas nem suportam completamente o aspecto interactivo.

Este projecto visa o desenvolvimento de uma ferramenta web interativa, em Português, para ajudar de forma assistida e intuitiva a compreender os conceitos e algoritmos em causa, vendo-os a funcionar passo-a-passo, através de exemplos típicos pré-carregados ou construídos pelo utilizador (um aspecto original da nossa plataforma). A ferramenta deve, por isso, permitir criar e editar autómatos, bem como executar os algoritmos clássicos relevantes, como aceitação de palavras, conversões entre modelos, etc.
Pretende-se, também, visualizar não só o processo de construção do autómato, como todos os passos de aplicação de dado algoritmo. 

Esta ferramenta usa o \emph{Framework} Ocsigen, pois este proporciona o desenvolvimento de ferramentas web completas e interactivas escritas em OCaml, uma linguagem funcional com um forte sistema de verificação de tipos e, por isso, perfeita para se obter uma página web sem erros. O Ocsigen foi escolhido também porque permite a criação de páginas dinâmicas com sistema de cliente-servidor único.

Este artigo apresenta a primeira fase do desenvolvimento do projecto, sendo já possível criar autómatos, aferir a natureza dos seus estados e verificar passo-a-passo (com \emph{undo}) a aceitação de uma palavra.

\begin{keywords}
Linguagens Formais, Autómatos, OCaml, Ocsigen, Ensino, Páginas Web Interactivas.
\end{keywords}
   
\end{abstract}

\newcommand{\novathesis}{\emph{novathesis}}
\newcommand{\novathesisclass}{\texttt{novathesis.cls}}

\ifshortversion 
    \section{Introdução}
\fi
\label{cha:introduction}
Dado o carácter matemático e formal dos tópicos abordados em matérias como linguagens e autómatos, o seu ensino e aprendizagem são exigentes e desafiantes. Vários estudos comprovam estas dificuldades e avaliam a utilidade de aplicações para estudar este tema \cite{Cogliati05, Sanders15, Pillay09, D'antoni15}. É importante dar apoio ao trabalho autónomo dos alunos com ferramentas interativas que permitam visualizar exemplos e fazer exercícios. No entanto, a maioria das aplicações são em Inglês e, por terem focos diferentes, nem sempre respondem a todas as necessidades. Enquanto umas são bastante completas, mas por serem \textit{Desktop}, nem sempre estão acessíveis, outras, são de fácil acesso através do \textit{browser}, embora estejam pouco desenvolvidas.

No contexto Português, no que diz respeito aos programas e bibliografias da maioria das disciplinas de Teoria da Computação (ou equivalentes) nas principais Universidades, verifica-se que o material é essencialmente teórico e, que apesar de aquele que é usado em sala de aula poder ser em português, a bibliografia é essencialmente em Inglês. Há, por isso, lugar para uma ferramenta interativa, em Português, que complemente o estudo teórico que já é feito hoje em dia nas aulas.

O objectivo deste projecto é desenvolver uma aplicação, em Português, que facilite o estudo da Teoria da Computação para alunos de Informática, disponível através de um \textit{browser}. A ideia é criar uma ferramenta que possa vir a suportar todos os tópicos dentro da Teoria da Computação, como autómatos finitos deterministas e não deterministas, autómatos de pilha, linguagens regulares, linguagens independentes de contexto, linguagens LL e todas as funcionalidades inerentes as estes tópicos, como conversões, minimizações e testes. Procura-se, ainda, criar uma ferramenta extensível que permita acrescentar funcionalidades, de forma fácil e eficaz. Pretende-se, também, que esta ferramenta esteja, em primeiro lugar, adaptada à disciplina lecionada na FCT-UNL, que venha a incluir exercícios avaliados de forma automática e com \textit{feedback} e que permita, ainda, aos alunos criar os seus próprios exercícios.

Esta plataforma está a ser desenvolvida em \textit{Ocsigen Framework}, ferramenta que permite a criação de sistemas web interativos, totalmente escrito em \textit{OCaml}. Ao tirar partido das características do \textit{OCaml}, é possível obter páginas web completamente funcionais e menos sujeitas a erros. O \textit{Ocsigen} facilita, ainda, a criação de ferramentas web, pois permite escrever cliente e servidor na mesma linguagem, facilitando a programação do sistema.

Neste documento, é apresentada a primeira versão da ferramenta e o seu desenvolvimento. Esta é, para já, uma página simples, mas com algumas funcionalidades importantes: visualização de autómatos,  (exemplos disponíveis na aplicação ou criados pelo utilizador), teste de aceitação de palavra (passo a passo ou de forma animada) e verificação da natureza dos estados. Esta ferramenta pode ser acedida em \url{http://ctp.di.fct.unl.pt/FACTOR/OFLAT} e o código está disponível para consulta em \url{https://bitbucket.org/rpmacedo/oflat/src/master/}.

\ifshortversion
\section{Trabalho Relacionado}
\fi
\label{cha:state_art}

\ifshortversion
\subsection{Ferramentas de visualização existentes}
\fi
Desde cedo se percebeu que as dificuldades de aprendizagem, e mesmo de ensino, no Estudo das Linguagens Formais e Teoria de Autómatos era um problema recorrente. E que, por ser um tema que gira em torno de processos e máquinas abstractas, a melhor forma de compreender esta problemática seria através de ferramentas pedagógicas. O desenvolvimento destas ferramentas tem sido feito desde o inicio da década de 60 \cite{Chakraborty11}. O artigo \textit{Fifty Years of Automata Simulation: A Review} \cite{Chakraborty11} defende também que de já existirem muitas ferramentas a comunidade cientifica continua a receber novas pois cada uma é diferente, cada uma tem os seus próprios princípios e muitas das vezes novas utilidades. Para além disso cada ferramenta é influenciada pelas ferramentas de desenvolvimento disponíveis no momento.

Existem desde ferramentas textuais, como por exemplo \cite{Dias05}, que permite criar e testar autómatos através de texto ou código, sem que estes sejam visíveis graficamente, e ferramentas baseadas na visualização gráfica. Algumas destas serão mais à frente analisadas, com maior profundidade, por terem semelhanças com o tema deste projecto.

No trabalho de pesquisa efectuado, foram encontrados muitos exemplos de aplicações que, de alguma forma, visam colmatar as dificuldades mencionadas pelo recurso a soluções diferenciadas. A título de exemplo, referem-se as seguintes ferramentas: \textit{FSM Simulator} \cite{Grinder03} um programa Java que possibilita fazer simulações com autómatos finitos; \textit{Language Emulator} \cite{Vieira04} uma ferramenta que permite trabalhar com diferentes conceitos dentro da Teoria de Autómatos e que tem sido usada por estudantes na Universidade de Minas Gerais no Brasil; \textit{jFAST} \cite{Thomas06} um software gráfico que permite o estudo de máquinas de estado finitas; \textit{RegeXeX} \cite{Brown07} um sistema interactivo para estudar Expressões Regulares; \textit{Forlan} \cite{Stoughton08} ferramenta embebida na linguagem Standard ML que permite a experimentação de linguagens formais e autómatos.

É ainda importante referir a ferramenta descrita por \textit{Coffin et. al} em \cite{Coffin63} por ser, provavelmente, a primeira a ser desenvolvida na área. Muitas outras são referidas no artigo \textit{Fifty Years of Automata Simulation} \cite{Chakraborty11}.

Pode-se falar ainda de bibliotecas desenvolvidas para se Linguagens Formais e Autómatos como Awali \cite{awali} (evolução de Vaucanson \cite{lombardy03,lombardy05}  e Vaucanson 2 \cite{lombardy13}) e Grail \cite{Raymond94},escritas em c++; FAdo, escrita em Python. Awali e FAdo são também e exemplo de bibliotecas que estão a evoluir para aplicações gráficas.

Convém também referir aplicações como \textit{TANGO} \cite{Stasko90}, \textit{JAWAA} \cite{Pierson98}, \textit{GUESS} \cite{Adar06} e \textit{VisualAlgo} \cite{visualgo}, por terem o objectivo de animar algoritmos ou estruturas de dados, estando relacionados com este projecto devido à sua forte componente interactiva.

Pelo que ficou dito, compreende-se que existem muitas ferramentas que poderiam ser aqui mencionadas. Decidiu-se, contudo, desenvolver um pouco mais aquelas que de alguma forma se destacam por diferentes especificidades como JFlap por ser, provavelmente, a ferramenta mais completa disponível; Automaton simulator por ser uma ferramenta web que permite estudar AF, permitindo verificar a aceitação de frases; FSM simulator, Regular Expression Gym e FSM2Regex por serem também aplicações web e permitirem o estudo de Autómatos Finitos e Expressões Regulares e as suas conversões; Automata Tutor v2.0 por ter um sistema de avaliação e feedback; e AutoMate por ser uma ferramenta com objectivos semelhantes ao deste projecto e que se encontra em desenvolvimento ao mesmo tempo.

\ifshortversion
\subsubsection{JFlap}
\fi
\label{sec:jflap}

\begin{figure}[htbp]
  \centering
  \subcaptionbox{Editor de Autómatos \label{fig:JFLAP1}}%
    {\includegraphics[width=0.5\linewidth]{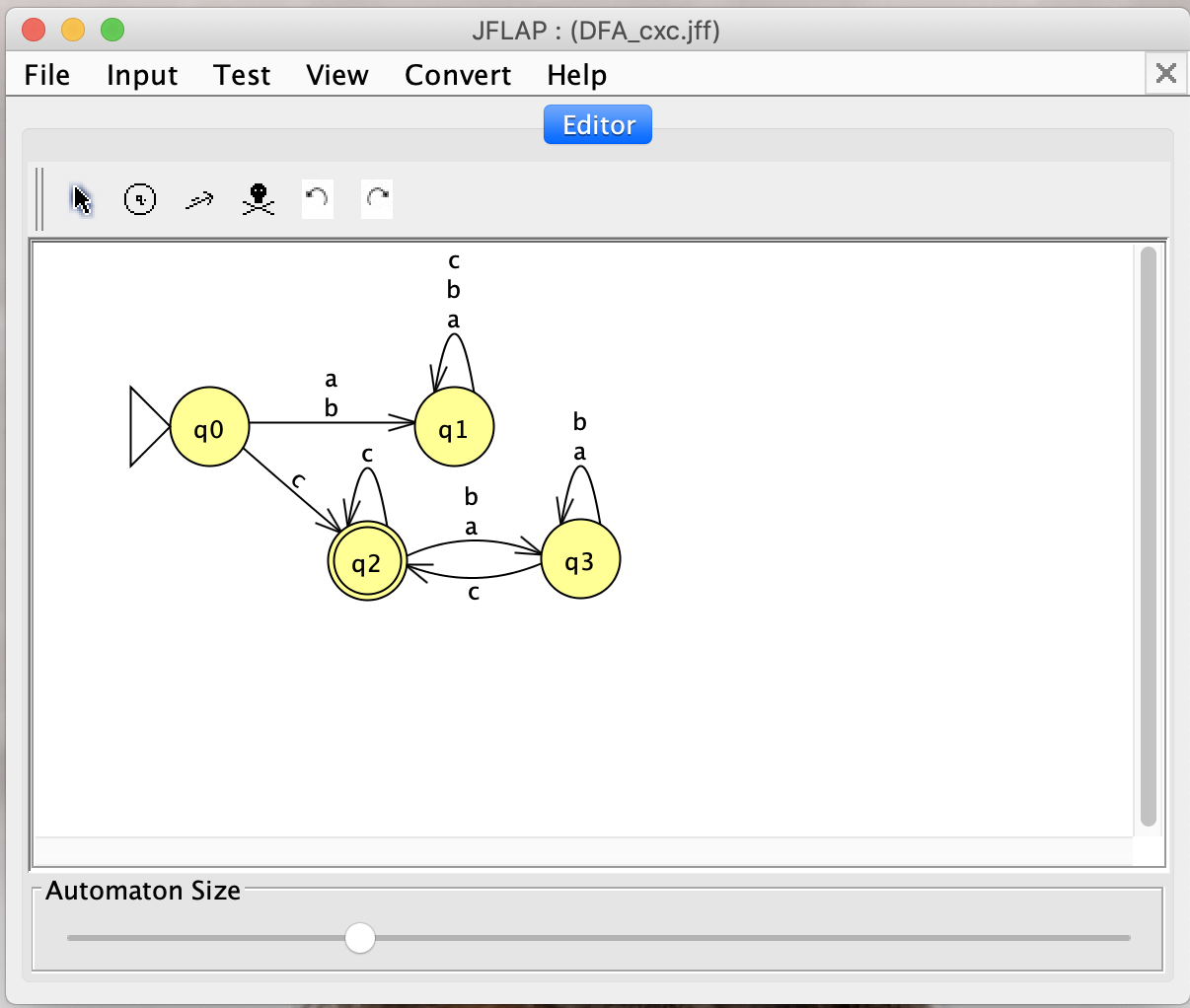}}%
  \subcaptionbox{Página para testar passo a passo o autómato \label{fig:JFLAP2}}%
    {\includegraphics[width=0.5\linewidth]{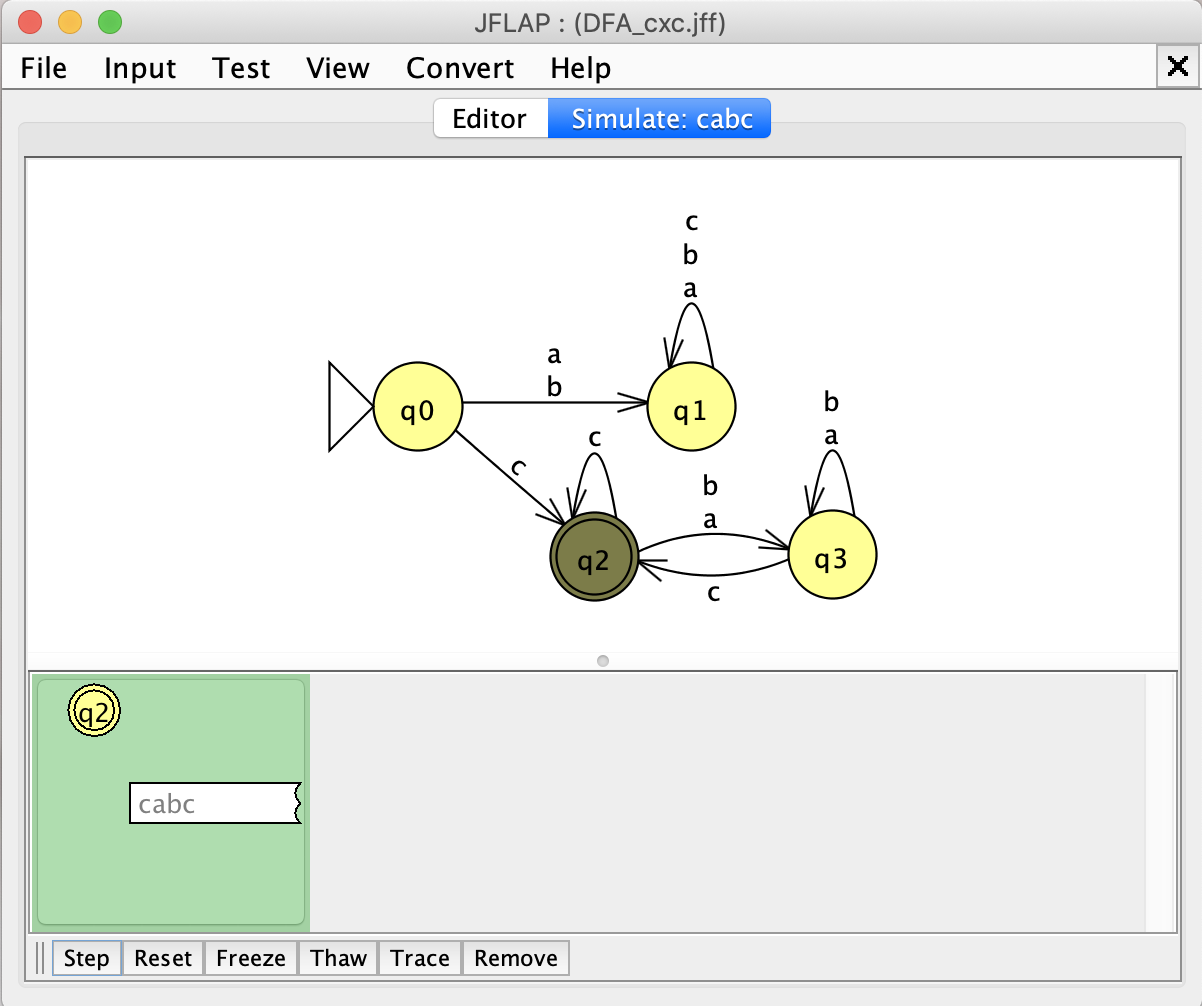}}%
  \caption{JFLAP exemplos \cite{Automate}}
  \label{fig:JFLAP}
\end{figure}

\cite{Jflap} é uma ferramenta \textit{desktop} que se encontra em desenvolvimento desde 1990. Destaca-se por ser uma das mais completas para o estudo de Linguagens Formais e Teoria de Autómatos. 

É fruto do trabalho de Susan H. Rodger e de alguns dos seus alunos que, ao longo do tempo, foram desenvolvendo novas funcionalidades \cite{JflapHistory, JflapHistoryPDF}. Apesar de ter sido inicialmente escrita em c++ e \textit{x windows}, foi mais tarde reescrita em Java e \textit{swing} de forma a melhorar a interface gráfica.

O código fonte está disponível na página web \cite{Jflap} e no GitHub, permitindo modificações por qualquer utilizador. 

Este software é complementado por uma página web que contém explicações e resolução de exercícios e por um livro guia para utilização da aplicação, mas que se encontra desatualizado.

JFLAP é usado há mais de 20 anos em diversas universidades do mundo, tendo já sido testado o seu impacto positivo \cite{rodger98, Merceron07, Rodger09}.

Em termos de funcionalidades, é possível trabalhar de forma interativa com Autómatos finitos (converter AFNs em AFDs, AFNs em expressões ou gramáticas regulares, minimizar AFDs, testar se aceitam palavras e visualizar o processo de aceitação passo a passo); Máquinas de Mealy; Máquinas de Moore; Autómatos de Pilha (criação a partir de linguagens livres de contexto e vice-versa); Três tipos de Máquinas de Turing (de uma fita, de múltiplas fitas, através de blocos); Gramáticas; Sistema-L, Expressões Regulares (criar AFDs, AFNs, gramáticas regulares e expressões regulares); Lema da bombagem para Linguagens Regulares; Lema da bombagem para linguagens livre de contexto. 

Apesar de todos os seus pontos positivos, \textit{JFLAP} é uma aplicação desktop, o que significa que nem sempre está acessível, é necessário estar no computador e descarregar o software para se conseguir utilizar. Muito embora as funcionalidades tenham evoluído, o design encontra-se um pouco datado e a sua utilização nem sempre é intuitiva. É também uma aplicação não muito fácil de utilizar sendo que para os capítulos mais complexos da Teoria da Computação é necessário ler o livro de instruções para perceber como utilizar a aplicação.

\begin{figure}[htbp]
    \centering
    \includegraphics [width=1.0\linewidth] {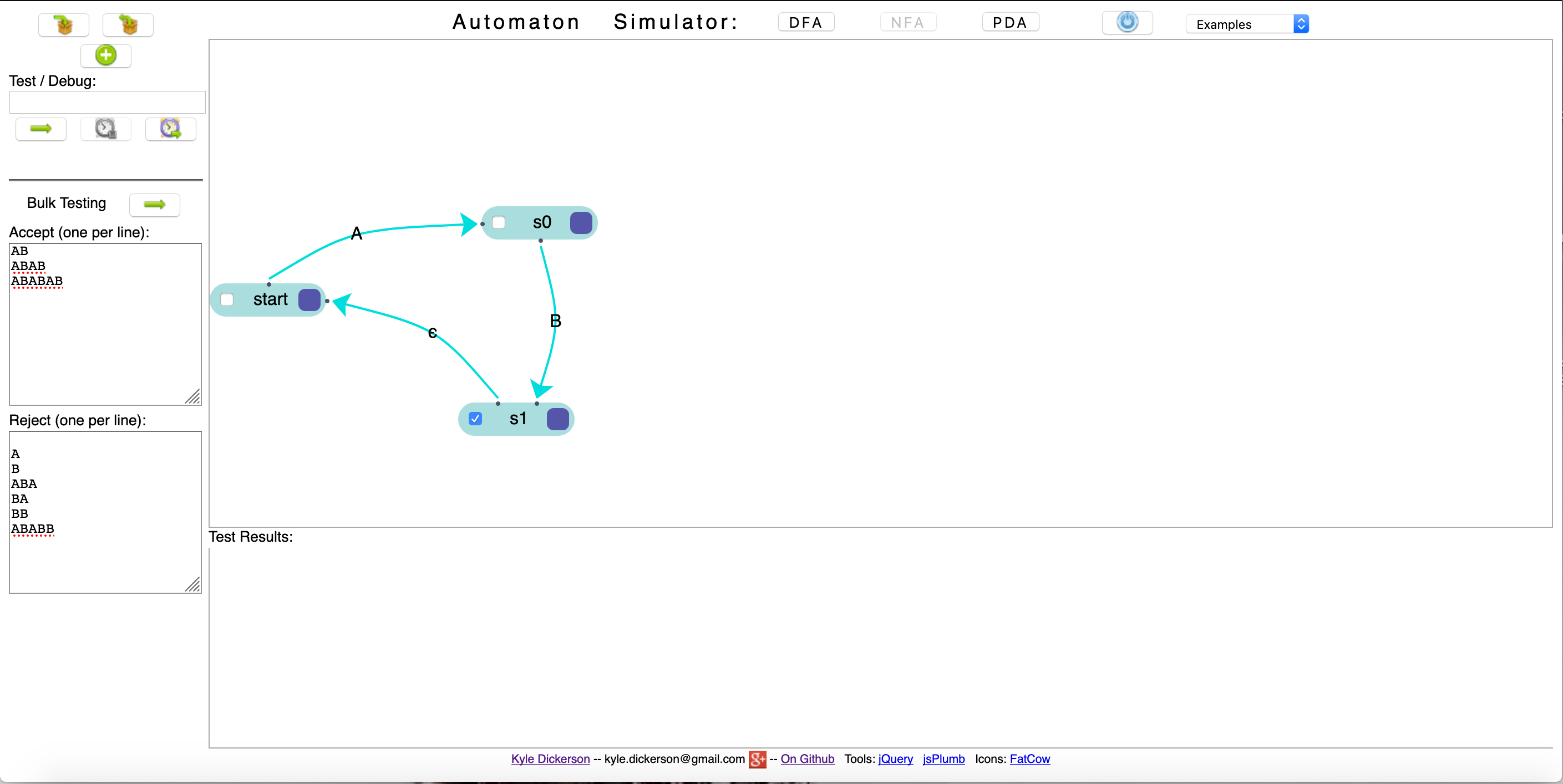}
  \caption{Página Automaton Simulator \cite{Automaton} com exemplo de frases para serem aceite e para serem rejeitadas}
  \label{fig:AutoSim}
\end{figure}

\ifshortversion
\subsubsection{Automaton Simulator}
\fi
\label{sec:automaton_sim}
\cite{Automaton} é uma ferramenta web muito simples, constituída por uma só página web (figura \ref{fig:AutoSim}). 

Nesta página é possível desenhar graficamente três diferentes tipos de autómatos - autómatos finitos deterministas, autómatos finitos não deterministas e autómatos de pilha. Não é possível gerar autómatos a partir de uma expressão regular, assim como também não é possível fazer a conversão de NFA em DFA.

Após a criação do autómato é possível testar a aceitação ou a rejeição de frases, bem como o reconhecimento passo a passo de uma frase pelo autómato. Contudo, só permite avançar para o passo seguinte e nunca voltar para trás.

Apesar da sua simplicidade, esta página resulta pouco intuitiva, uma vez que é desenhada à base de ícones, sem conter qualquer tipo de explicação. Além disso, contém poucas funcionalidades.

\ifshortversion
\subsubsection{FSM simulator \cite{FSM}, Regular Expressions Gym \cite{Gym}, FSM2Regex \cite{Regex}}
\fi
\label{sec:fsm}
são três ferramentas complementares que permitem o estudo de expressões regulares e teoria de autómatos. Cada uma destas ferramentas é uma página web desenvolvida, desde 2012, por Ivan Zuzak, Web Engineer e antigo professor na universidade de Zagreb, e Vedrana Jankovic, Engenheira de Software na Google. Cada página foi desenvolvida em \textit{Noam} - uma biblioteca \textit{JavaScript} que permite trabalhar com máquinas de estado finitas, gramáticas e expressões regulares -, \textit{Bootstrap, Viz.js} e \textit{jQuery}. O código fonte está disponível no GitHub, sob a licença Apache v2.0.

\begin{figure}[htbp]
    \centering
    \includegraphics [height=2.8in]{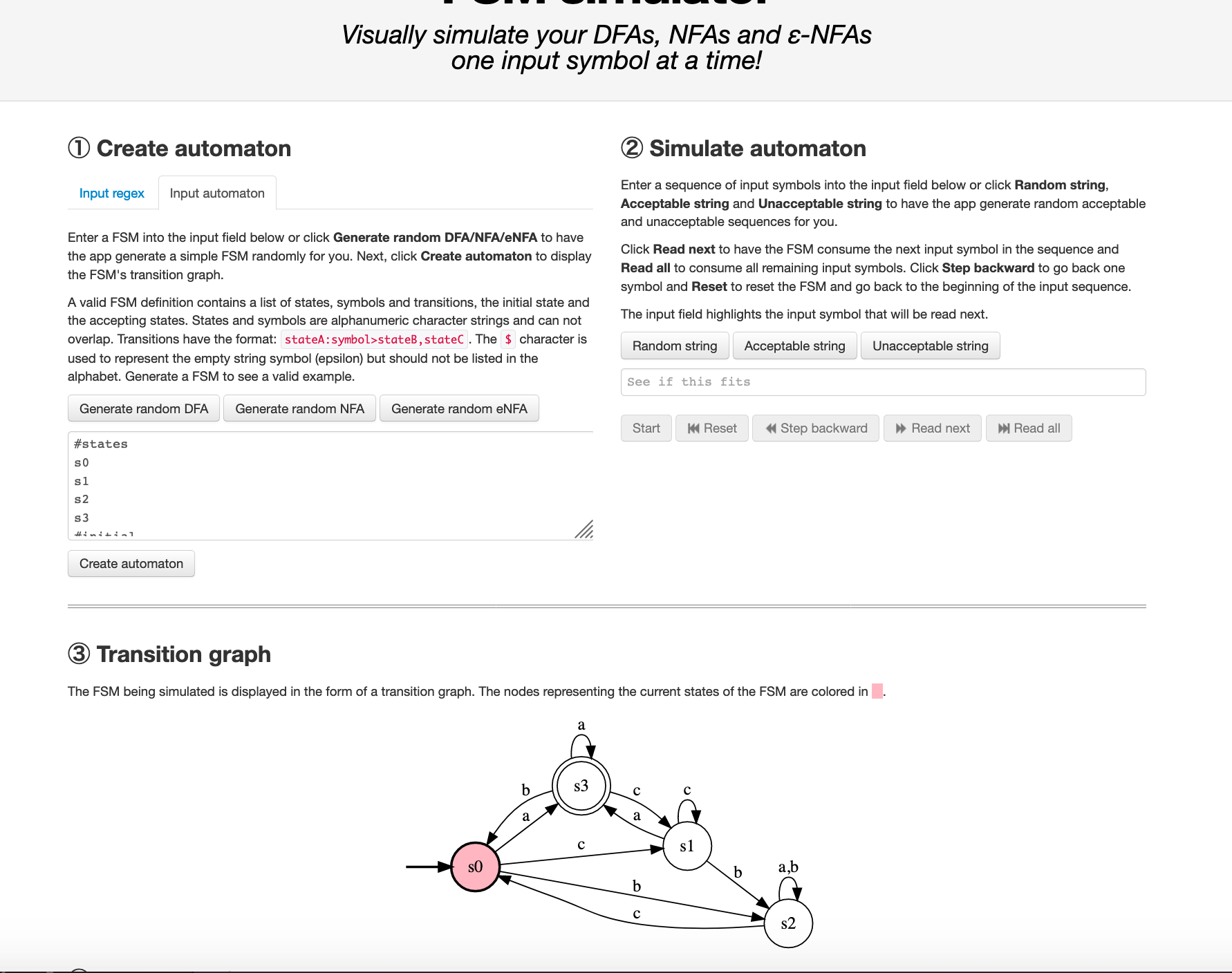}
  \caption{Página FSM Simulator \cite{FSM} com exemplo de um autómato}
  \label{fig:FSM}
\end{figure} 

\textit{FSM Simulator} (figura \ref{fig:FSM}) é utilizada para a criação e teste de autómatos. Os autómatos podem ser gerados através de expressões regulares ou através de texto, não sendo, no entanto, possível criá-los graficamente. É possível visualizar o processo de reconhecimento de uma frase pelo autómato, passo a passo, podendo o utilizador avançar ou retroceder, sempre que necessário.

\begin{figure}[htbp]
    \centering
    \includegraphics [height=2.8in] {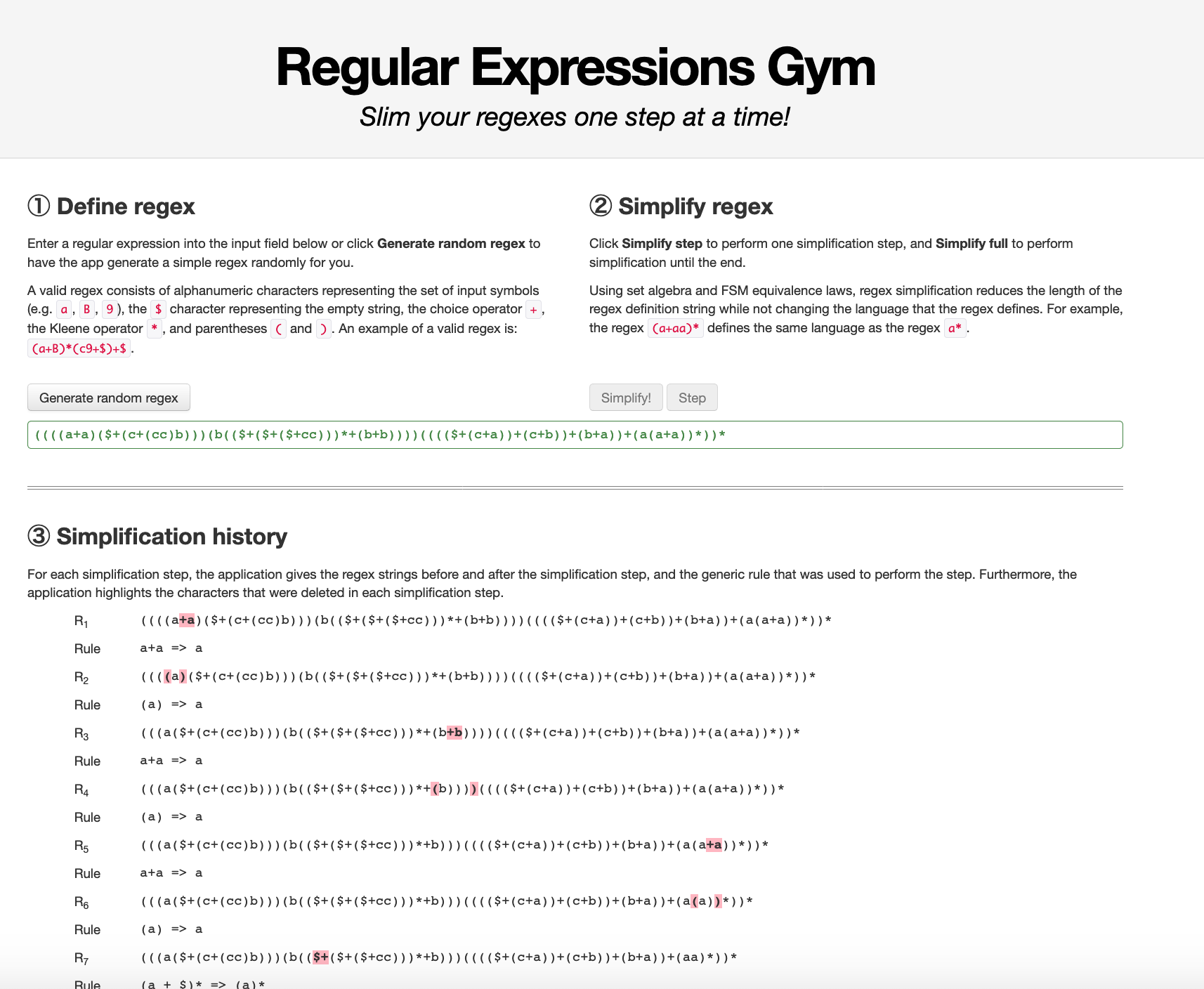}
  \caption{Página Regular Expression Gym \cite{Gym} com exemplo de uma expressão a ser reduzida}
  \label{fig:Gym}
\end{figure} 

\textit{Regular Expressions Gym} (figura \ref{fig:Gym}) é uma pagina web muito simples que permite visualizar a simplificação de uma expressão regular. O utilizador pode ver toda a simplificação de uma só vez ou pode escolher vê-la passo a passo até estar terminada. Para cada passo da simplificação, em qualquer uma das opções, é indicada a regra utilizada.

\begin{figure}[htbp]
    \centering
    \includegraphics [height=2.5in] {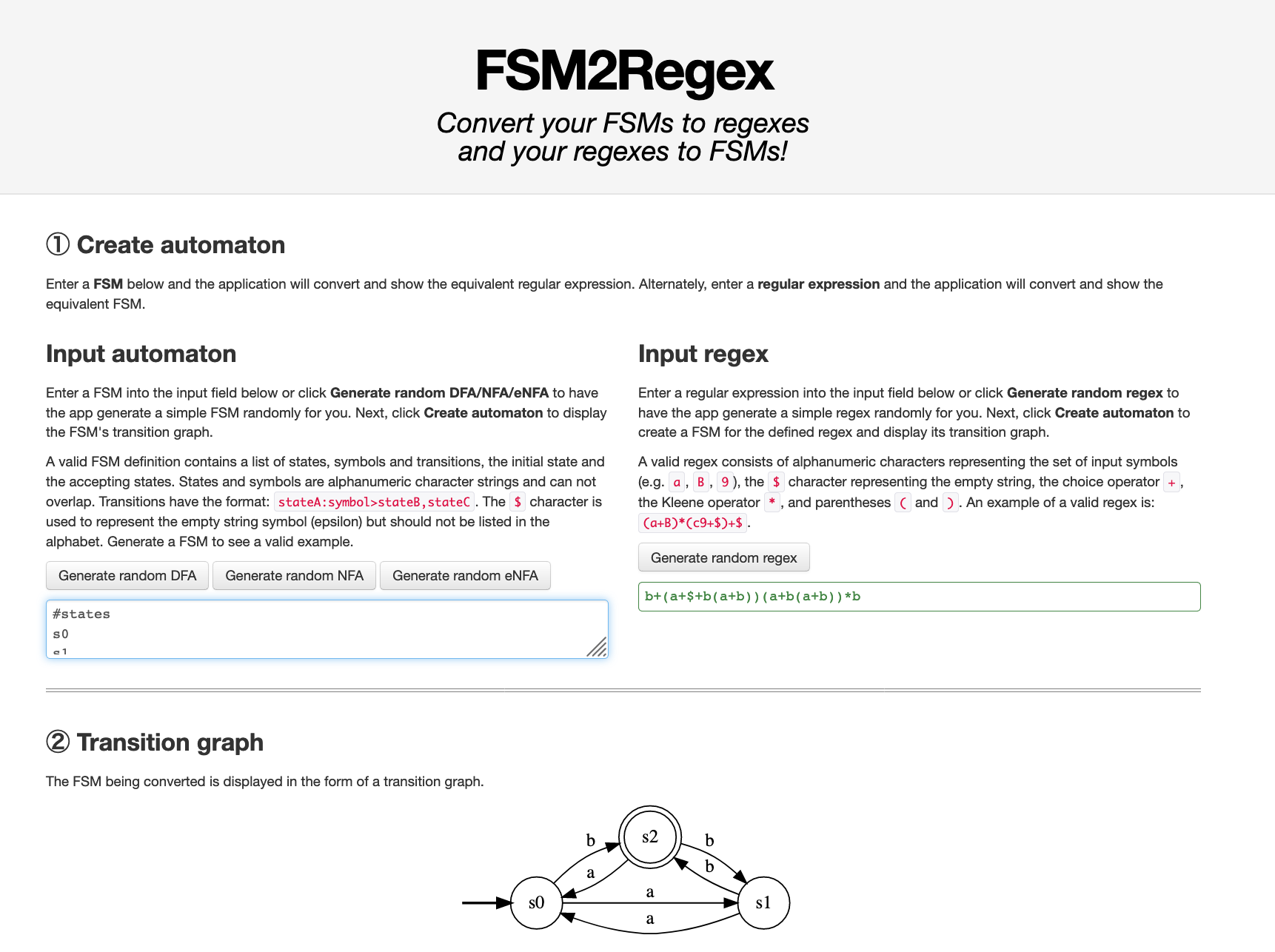}
  \caption{Página FSM2Regex \cite{Regex} de um autómato e da sua correspondente expressão regular}
  \label{fig:Regex}
\end{figure}

\textit{FSM2Regex} (figura \ref{fig:Regex}) é utilizada para fazer a conversão de uma expressão regular num autómato e vice versa, não permitindo nenhum outro tipo de interacção.

As três páginas abordadas correspondem a ferramentas bastante simples e intuitivas, mas seriam, provavelmente, mais proveitosas se as funcionalidades de cada uma delas fossem integradas numa só página. Dessa forma seria também mais fácil o desenvolvimento de novas funcionalidades.

\ifshortversion
\subsubsection{Automata Tutor v2.0}
\fi
\label{sec:automata_tutor}
\ifshortversion
\cite{Automatatutor, dantoni15} é uma ferramenta web que se destaca por fornecer um sistema de avaliação de exercícios de autómatos finitos e expressões regulares, facilitando o trabalho dos professores. Esta ferramenta passou por três fases de desenvolvimento e vários testes por utilizadores \cite{dantoni15, D'antoni15}, com o objectivo de melhorar a aplicação. 

A página permite o registo e entrada no sistema com dois tipos de perfis: o de professor, ao qual é dada a possibilidade de criar um curso com exercícios próprios e visualizar as notas no final do curso; e o de aluno, que se pode inscrever num determinado curso ou então fazer os exercícios disponíveis na própria página. O foco principal desta ferramenta é a resolução de exercícios de criação de expressões regulares e autómatos finitos correspondentes a uma frase dada em Inglês. Quando um aluno submete uma solução, recebe como resposta, além da nota, alguns comentários, que lhe permite proceder a correcções, caso seja necessário. Apesar de ser uma aplicação muito completa quanto à avaliação de exercícios, esta foca-se somente da entrega de feedback, não dando liberdade de experimentar a resolução de forma autónoma para compreender porque está certo ou errado (como por exemplo a verificação passo a passo da aceitação da palavra).
\fi

\begin{figure}[htbp]
\centering
    {\includegraphics[width=0.7\linewidth]{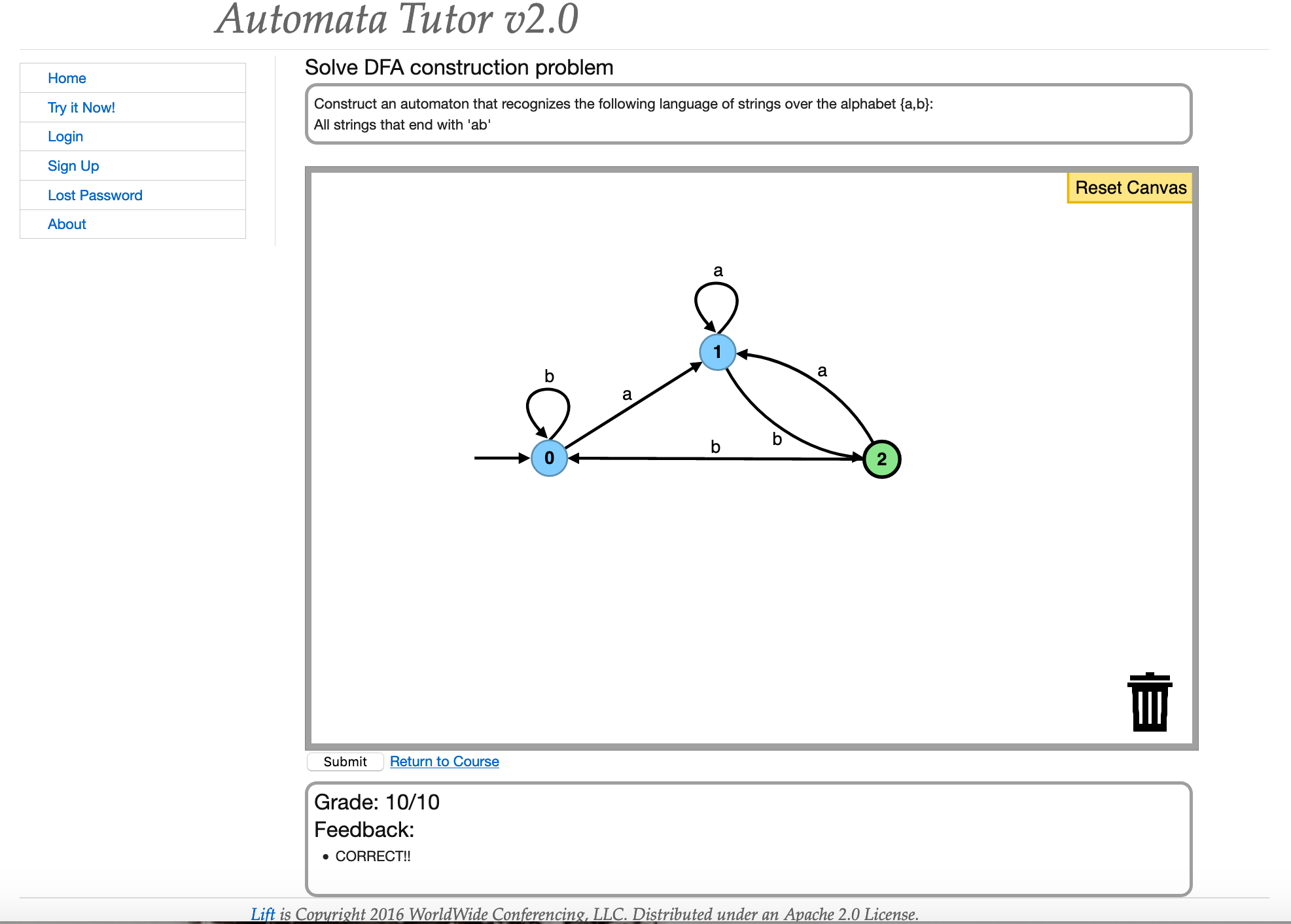}}%
  \caption{Exemplo da resolução de um exercicio DFA no AutomataTutor \cite{Automatatutor}}
  \label{fig:Tutor}
\end{figure}


\ifshortversion
\subsubsection{AutoMate} 
\fi
\label{sec:automate}
\cite{Automate} é uma ferramenta web muito recente, lançada em 2019 e encontrando-se ainda numa fase muito inicial. É, no entanto, relevante falar dela, uma vez que se encontra simultaneamente em desenvolvimento com a proposta deste projecto. Tem como finalidade apoiar o estudo de alunos de Informática, ou seja, pretende ajudar a resolver exercícios sobre autómatos finitos e expressões regulares. 

\begin{figure}[htbp]
  \centering
  \subcaptionbox{Página para ver e resolver exercícios \label{fig:exercisePage}}%
    {\includegraphics[width=0.5\linewidth]{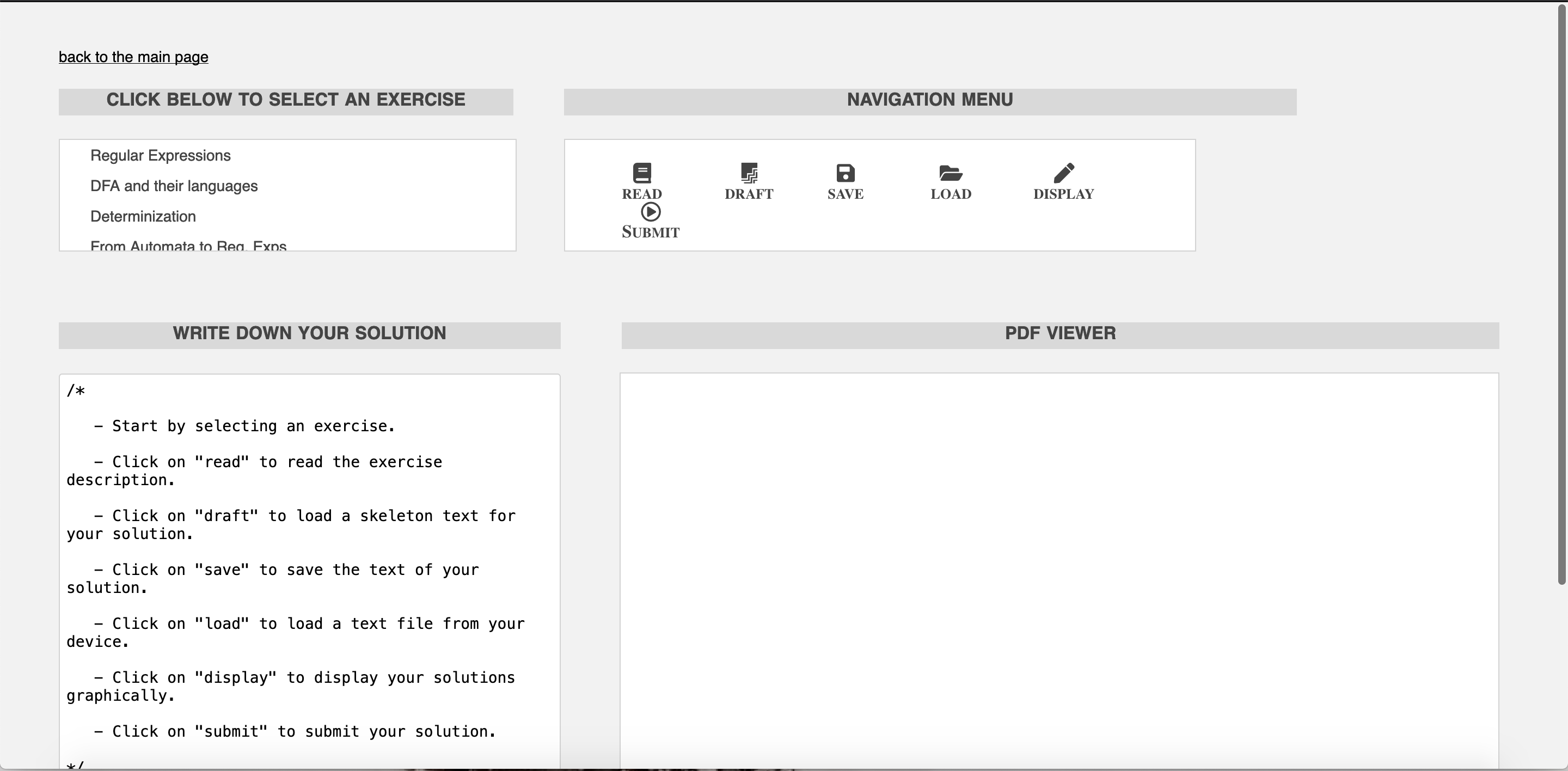}}%
  \subcaptionbox{Página para desenhar autómatos \label{fig:automataPage}}%
    {\includegraphics[width=0.5\linewidth]{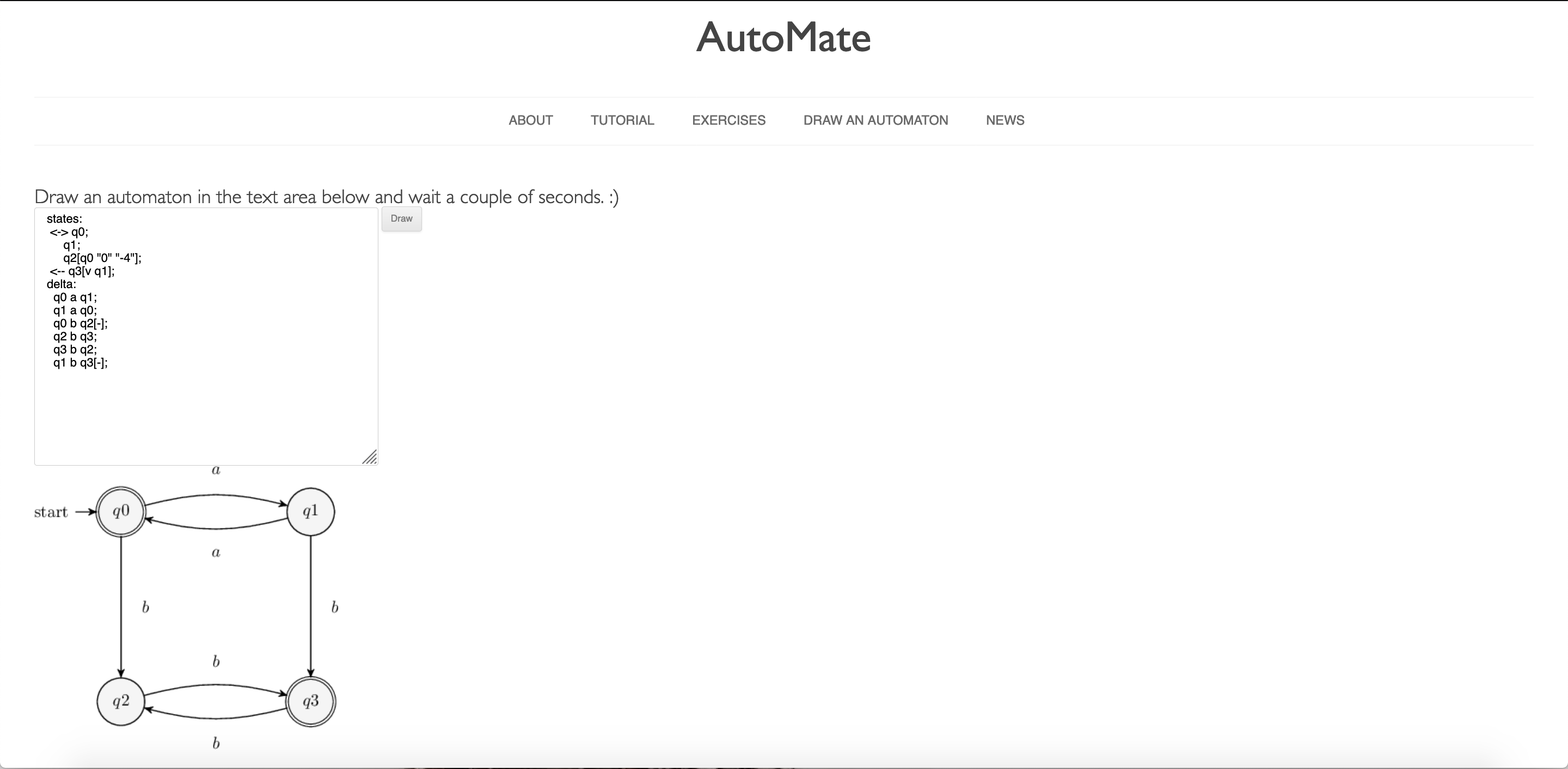}}%
  \caption{Exemplo de duas páginas da Ferramenta Web Automate  \cite{Automate}}
  \label{fig:automate}
\end{figure}

Em termos de funcionalidades esta ferramenta é, para já, bastante simples, centrando-se principalmente na verificação e correcção de exercícios (figura \ref{fig:exercisePage}). Todo o processo de realização de exercícios é feito através de texto, não sendo possível a criação de autómatos graficamente.

A ferramenta permite realizar exercícios de diferentes temas dentro da Teoria da Computação - Expressões Regulares, criação de DFAs, transformação NFAs em DFAs, passagem de DFAs para expressões regulares. Após a entrega da resolução dos exercícios é devolvido um \textit{feedback} automático, em formato pdf, relevante para que o aluno.

Esta ferramenta contém ainda uma página onde se pode desenhar um autómato (ver figura \ref{fig:automataPage}), através de texto, de forma a que seja possível visualizá-lo através de uma imagem. Não é no entanto possível a verificação da aceitação de uma palavra no autómato criado.

Em termos de design esta ferramenta é ainda muito simples e rudimentar (figura \ref{fig:automate}, sendo baseada em caixas de inserção de informação.

\bigskip

Todas as ferramentas mencionadas têm características que, de alguma forma, são comuns ao projecto em desenvolvimento mas não foi encontrada nenhuma que permitisse aos alunos criar os seus próprios exercícios de forma a receber feedback da solução.

\ifshortversion
\subsection{Ambientes de desenvolvimento para a Web}
\fi
Tipicamente, uma ferramenta web pressupõe dois componentes: cliente e servidor. Estes são normalmente desenvolvidos em linguagens diferentes e, por tal, para que os dados possam ser partilhados entre ambos é também necessário escrevê-los num formato pré-estabelecido.

O lado do cliente é a parte da aplicação web com a qual o utilizador interage, sendo que a cada interação é enviado um pedido ao servidor, que responde com a execução de uma ação ou uma informações da base de dados. Este é maioritariamente desenvolvido pelo recurso a três linguagens: HTML, CSS, \textit{JavaScript}.

O lado do servidor é a parte da aplicação web que indica como esta funciona. O servidor recebe pedidos do cliente e retorna a informação pedida ou a ação que pode ser realizada. O servidor pode ser escrito em diferentes tipos de linguagens como C, C++, C\#, PHP, \textit{Python}, \textit{Java} ou até \textit{JavaScript}.

Para que o cliente possa comunicar com o servidor é utilizado um protocolo de comunicação, como HTTP, e a mensagem deve estar num formato convencionado, sendo os mais comuns HTML, XML ou JSON.

Com a necessidade de criação de páginas web mais dinâmicas, e devido ao facto de ser necessária a aprendizagem de muitas linguagens e componentes, começaram a surgir diferentes \textit{frameworks} e bibliotecas com o objectivo de facilitar o trabalho do programador. Alguns visam facilitar o desenvolvimento do design da página, como o \textit{Bootstrap}, cujo o objectivo é criar páginas web \textit{responsive}. Outros pretendem facilitar a criação de aplicações web de uma só página, isto é, reactivas. Temos, como exemplos, \textit{AngularJS}, um framework que estende a linguagem \textit{HTML}; e \textit{React}, uma biblioteca \textit{JavaScript}. De referir ainda \textit{jQuery} uma biblioteca \textit{JavaScript} que pretende simplificar a escrita das \textit{queries} do mesmo e \textit{frameworks} que visam facilitar o desenvolvimento de servidores com muitos acessos à base de dados e que pressupõem reutilização de código, como \textit{Django} e \textit{Ruby on Rails}. 

Apesar do surgimento de tantos \textit{frameworks} e bibliotecas, o programador, para criar páginas web, necessita sempre de saber pelo menos \textit{HTML, CSS, JavaScript} (ou um correspondente) e uma linguagem para o servidor.

\label{cha:trabalho}

\section{Ocsigen Framework}
\textit{Ocsigen Framework} é uma ferramenta muito completa que se destaca principalmente por permitir a criação de páginas web interativas, escritas totalmente em \textit{OCaml}. Este \textit{Famework} surge da ideia de que a programação funcional é uma solução elegante para alguns problemas de interação nas páginas web \cite{Balat09}. Vem tentar responder aos novos desafios das páginas web, isto é, à necessidade de estas se comportarem cada vez mais como aplicações \cite{Balat06}. Uma das grandes vantagens é compilar o código \textit{OCaml} do cliente para \textit{JavaScript}, o que permite trabalhar em conjunto com esta linguagem e, assim, utilizar um amplo número de bibliotecas, que de outra forma não estariam disponíveis.

\subsection{Descrição das componentes}
\textit{Ocsigen Framework} é na realidade um conjunto de diferentes componentes, o que traduz a complexidade desta ferramenta.

\textbf{Eliom} é o componente principal do \textit{Ocsigen}, é uma extensão do \textit{OCaml} para programação sem camadas (\textit{Tierless}) \cite{BalatVincente16}. \textit{Eliom} pretende ser um novo estilo de programação que se enquadra nas necessidades das aplicações web modernas melhor do que as linguagens de programação usuais (desenhadas há muitos anos, para páginas muito mais estáticas \cite{ocsigen}). O seu grande objectivo é permitir desenvolver uma aplicação distribuída, totalmente em \textit{OCaml} e como um só programa \cite{ocsigen}, ou seja, não haver separação entre cliente e servidor. Para que isto seja possível existe uma sintaxe especial para distinguir os dois. O cliente pode aceder facilmente a variáveis do servidor, pois o sistema de suporte implementa esse mecanismo de forma transparente. Outra importante vantagem resulta do uso da tipificação estática do Ocaml, possibilitando verificar erros e bugs no momento da compilação, havendo a certeza de que se obtêm páginas web corretas, sem problemas nos links ou na comunicação cliente-servidor. Além disso, o \textit{Eliom} resolve automaticamente problemas de segurança frequentes em páginas web. O \textit{Eliom} tem ainda, por base o pressuposto de que se escreve código complexo em poucas linhas. As aplicações desenvolvidas nesta ferramenta correm em qualquer \textit{browser} ou dispositivo móvel, não havendo necessidade de customização.

\textbf{Js\_of\_ocaml} é o compilador de \textit{OCaml bytecode} para \textit{JavaScript}. É o componente do \textit{Ocsigen} que permite correr a aplicação escrita em \textit{OCaml} em ambientes \textit{JavaScript} como os \textit{browsers} \cite{ocsigen}  \cite{vouillon-balat13}. Possibilita a integração de código \textit{JavaScript} no programa \textit{OCaml}.

\textbf{Lwt} é a biblioteca de \textit{threads} cooperativa para \textit{OCaml} que permite lidar com problemas de concorrência de dados e de \textit{deadlocks}. É a forma standard de construir aplicações concorrentes. Permite fazer pedidos de forma assíncrona, através de promessas. Estas são simplesmente referências que vão ser preenchidas assincronamente, aquando da chegada da resposta.

\textbf{Tyxml} é  a biblioteca que permite a construção de documentos HTML estaticamente corretos. Tyxml providencia um conjunto de combinadores, que utilizam o sistema de tipos do OCaml para confirmar a validade do documento gerado.

\textbf{Ocsigen-start} é a biblioteca e \textit{template} de uma aplicação \textit{Eliom} com muitos componentes típicos das páginas web, que visa facilitar a contrução de aplicações web interativas.

\textbf{Ocsigen-toolkit} é a biblioteca de \textit{widgets} que, tal como o \textit{Ocsigen-start}, visa facilitar o desenvolvimento rápido de aplicações web interativas.

\section{Animação e visualização de Autómatos}
O objectivo deste projecto é a criação de uma ferramenta web que venha a permitir a alunos de informática estudar os vários temas dentro da Teoria de Computação de forma intuitiva e eficaz. Pretende-se que esta ferramenta permita ao aluno trabalhar com Autómatos finitos (minimização, conversão de autómatos não deterministas em deterministas, minimização, conversão em expressões regulares, verificação de aceitação de palavras e geração de palavras  de tamanho x), Expressões regulares (simplificação e conversão em AFN), linguagens não regulares, linguagens LL, linguagens independentes de contexto (lema da bombagem e conversão em autómatos de pilha), Autómatos de pilha (conversão em linguagens independentes de contexto) e máquinas de Turing. Para além disso pretende-se ter um sistema de resolução de exercícios com entrega de feedback no qual o aluno pode fazer upload dos seus próprios exercícios.

Este projecto encontra-se agora em desenvolvimento e nesta secção apresentamos e explicamos as funcionalidades disponíveis da primeira versão da aplicação, que pode ser acedida em \url{http://ctp.di.fct.unl.pt/FACTOR/OFLAT}. O código completo pode ser consultado em \url{https://bitbucket.org/rpmacedo/oflat/src/master/}.

Ao longo do processo de criação desta versão surgiram alguns desafios que fizeram a utilização do \textit{framework} um pouco complicada. Muito possivelmente devido à mudança de sintaxe do \textit{OCaml} (a sintaxe \textit{camlp4} foi descontinuada para dar lugar a \textit{ppx}), o \textit{Ocsigen} necessitou de ser modificado e nos exemplos de utilização ainda existem algumas inconsistências. Apesar de tudo, no final, conseguiu-se obter uma solução elegante e eficiente que comprova as características do \textit{framework}.

Como este \textit{framework} permite o trabalho conjunto entre \textit{OCaml} e \textit{JavaScript}, para a realização da parte gráfica optou-se pela utilização da biblioteca \textit{JavaScript} de visualização de grafos \textit{Cytoscape.js} \cite{Cytoscape}, por ser muito completa, com várias opções de manipulação e edição dos grafos. Todo o trabalho de verificação ou edição dos grafos é realizado em \textit{Eliom}, sendo a biblioteca utilizada unicamente para representar o grafo, ou seja, é feita uma separação entre o lado lógico e o lado gráfico da aplicação.

\subsection{Visão geral}
\begin{figure}[htbp]
\includegraphics[width=1.0\linewidth]{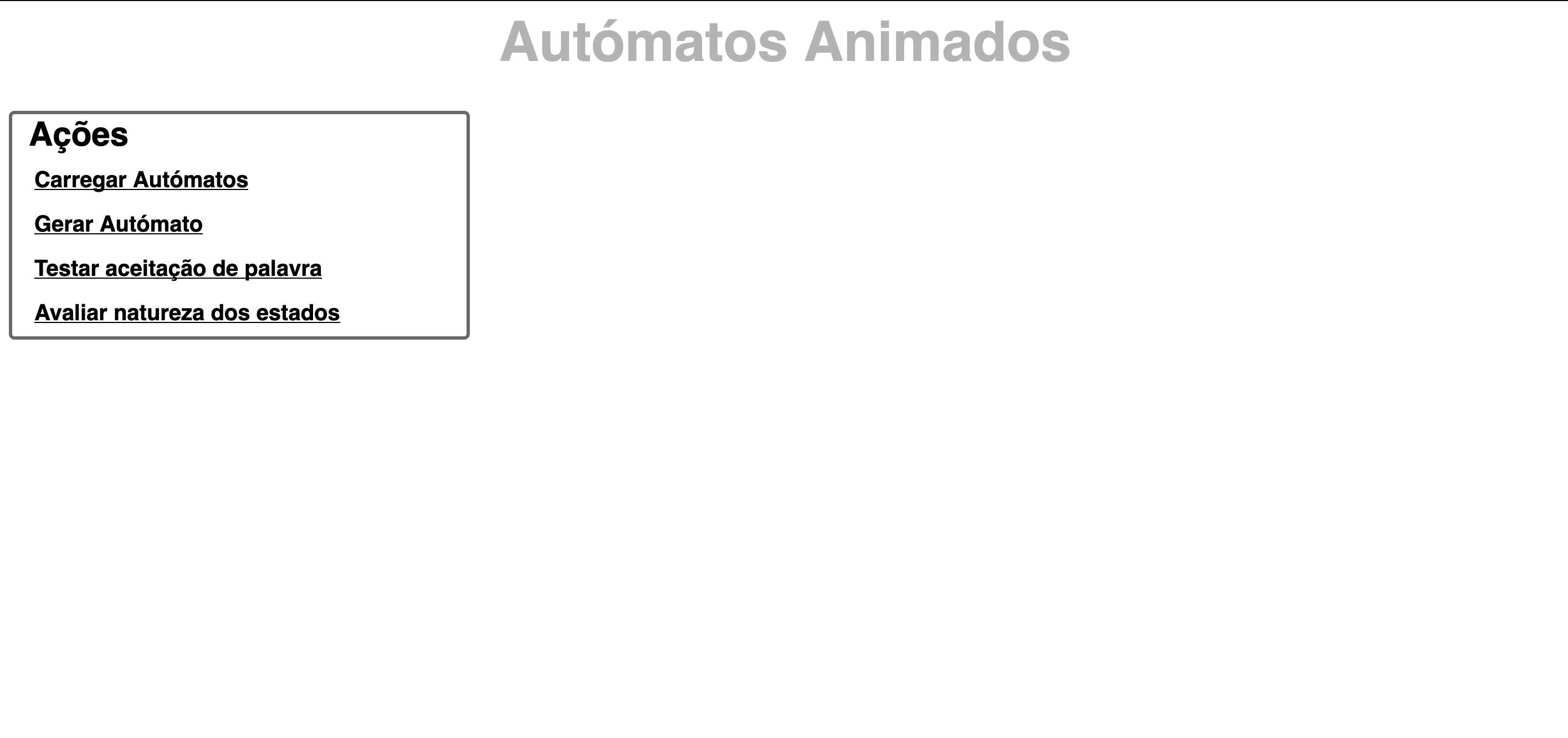}
\caption{Entrada da Aplicação}
\label{fig:entrada}
\end{figure}
Para se criar a página é necessária a criação de um módulo onde se regista a aplicação e de um serviço que indica o caminho ou \textit{link} para a página principal e, por fim, o registo do serviço no módulo criado, obtendo-se, assim, uma página principal. É no registo da página que se definem os elementos da mesma.

\begin{ocaml}
module Finalexample_app =
  Eliom_registration.App (
    struct
      let application_name = "finalexample"
      let global_data_path = None
    end)
let main_service = 
    Eliom_service.create
    ~path:(Eliom_service.Path [])
    ~meth:(Eliom_service.Get Eliom_parameter.unit)
    ()
let () =
  Finalexample_app.register
    ~service:main_service
    (fun () () ->
      let open Eliom_content.Html.D in
       Lwt.return
         (html
            (head (title (txt "Autómatos Animados"))
                [script ~a:[a_src script_uri1] (txt ""); 
                 script ~a:[a_src script_uri3] (txt "");
                 script ~a:[a_src script_uri5] (txt "");
                 script ~a:[a_src script_uri4] (txt "");
                 script ~a:[a_src script_uri6] (txt "");
                 script ~a:[a_src script_uri2] (txt ""); 
                 css_link ~uri: (make_uri (Eliom_service.static_dir ()) 
                                           ["codecss2.css"]) ();
                 script ~a:[a_src script_uri] (txt "");
            ])
            (body [ div [h1 [txt "Autómatos Animados"]];
                    div ~a:[a_id "inputBox"] 
                           [h2 [txt "Ações"];
                            mywidget "Carregar Autómatos" 
                                     (hiddenBox2 upload);
                            mywidget "Gerar Autómato" 
                                     (hiddenBox2 generate);
                            mywidget "Testar aceitação de palavra" 
                                     (hiddenBox2 verify);
                            mywidget "Avaliar natureza dos estados" 
                                     (hiddenBox2 evaluate);
                           ];
                    div ~a:[a_id "cy"] [];
                    ]
            )))    
\end{ocaml}

\begin{figure}[htbp]
  \centering
  \subcaptionbox{Menu principal \label{fig:menu1}}%
    {\includegraphics[width=0.5\linewidth]{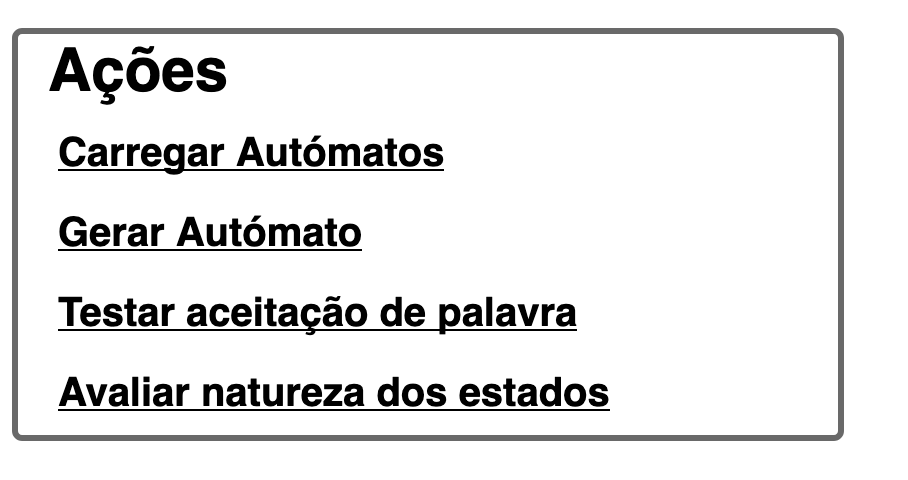}}%
  \subcaptionbox{Menu principal com opções \label{fig:menu2}}%
    {\includegraphics[width=0.5\linewidth]{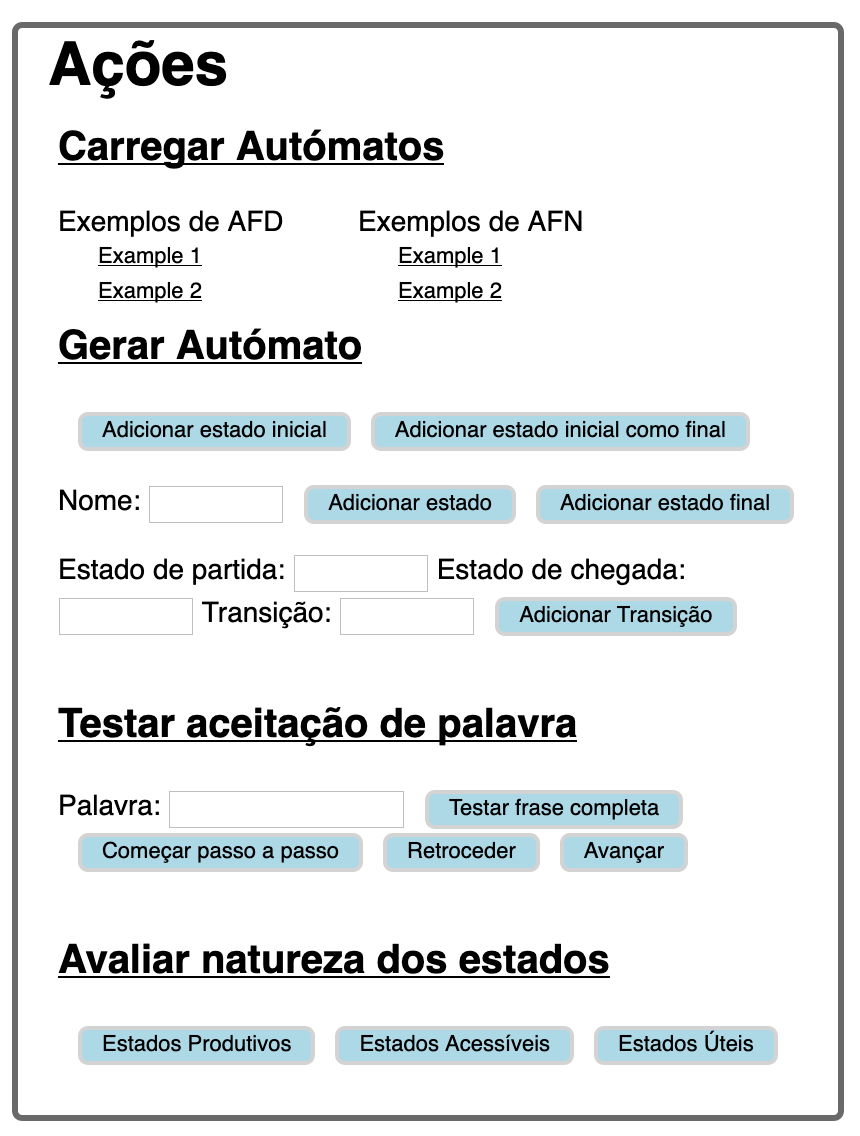}}%
  \caption{Vista do menu}
  \label{fig:menu}
\end{figure}

Quando se inicia a aplicação vê-se uma página simples (Figura \ref{fig:entrada}) que contém um menu do lado esquerdo e um espaço em branco do lado direito. Os autómatos, quando gerados, vão aparecer no lado direito do ecrã. No menu (Figuras \ref{fig:menu1}) são visíveis as opções disponíveis para se interagir com autómatos: carregar autómatos, gerar autómatos, testar aceitação de palavra e verificar a natureza dos estados. Ao clicar, em cada uma das opções principais, é possível visualizar as sub-opções disponíveis (Figura \ref{fig:menu2}). Cada chamada a mywidgets no código acima é a criação dos sub-menus disponiveis.

\subsection{Carregar Autómatos}
Na opção "Carregar Autómatos" são disponibilizados diferentes exemplos pré-definidos de autómatos. Para a resolução desta etapa foi necessário responder a diferentes questões, nomeadamente: Como aceder à biblioteca \textit{Cytoscape.js} programando em \textit{Ocsigen}? Como chamar funções do código \textit{JavaScript} em \textit{Ocsigen}? Como transformar um autómato formatado em \textit{OCaml} no grafo \textit{Javascript}? \smallskip

\textbf{Como aceder à biblioteca \textit{Cytoscape.js} programando em \textit{Ocsigen}?} Para responder a esta pergunta era necessário responder a duas subquestões: Como chamar \textit{scripts} ao iniciar a página (para se aceder a biblioteca durante a execução)? Como criar links? 

A pesquisa desenvolvida permitiu perceber que, no momento de registo do serviço principal, é possível chamar diferentes tipos de \textit{scripts} (visto no código da subsecção anterior) aquando do desenho da página. Era necessário ainda perceber como criar links para páginas externas. Após alguma pesquisa, encontrou-se alguns exemplos na página web da  ferramenta que possibilitaram  perceber como criar os links:
\begin{ocaml}
let script_uri1 =
  Eliom_content.Html.D.make_uri
  (Eliom_service.extern
     ~prefix:"https://unpkg.com"
     ~path:["cytoscape";"dist"]
     ~meth:(Eliom_service.Get Eliom_parameter.(suffix (all_suffix "suff")))())
    ["cytoscape.min.js"]
\end{ocaml}

\textbf{Como chamar funções do código \textit{JavaScript} em \textit{Ocsigen}?}
Responder a esta questão acabou por não ser tarefa fácil, pois a API encontrada na página web do framework apesar de ser muito completa contém pouca informação em  relação à comunicação entre código OCaml e código JavaScript. Com alguma pesquisa percebeu-se que tal como se chama o link da biblioteca, é possível chamar o link do código \textit{JavaScript}. Em seguida, para chamar as funções JavaScript é necessário usar outras duas funções \textit{OCaml}:
\begin{ocaml}
let js_eval s = Js_of_ocaml.Js.Unsafe.eval_string s
let js_run s = ignore (js_eval s)
\end{ocaml} 

Aquando da necessidade de aceder ao código \textit{JavaScript} chama-se a função \textit{js\_run} com o nome da função \textit{JavaScript} entre aspas.
\begin{ocaml}
    js_run ("makeNode1('"^f^"', '"^string_of_bool (test)^"')");
\end{ocaml}
\textit{js\_run}, por sua vez, chamará a função \textit{js\_eval}, como se pode ver pelo código acima. \smallskip

\textbf{Como transformar um autómato formatado em \textit{OCaml} no grafo \textit{JavaScript}?} Para se conseguir representar graficamente qualquer tipo de autómato sem que seja necessário este estar escrito tanto no código \textit{OCaml} como no \textit{JavaScript} foi preciso criar várias funções que fizessem a passagem da formatação \textit{OCaml} para o \textit{JavaScript}. Desta forma não há repetição de código, uma vez que os autómatos estão apenas representados no programa \textit{OCaml}.
\begin{ocaml}
let example1DFA = {initialState = "START";
                   transitions = [("START", 'a', "A"); ("A", 'b', "B"); ("B", 'a', "C"); ("C", 'b', "B"); ("C", 'a', "A")];
                   acceptStates = ["START"; "B"; "C"]}
\end{ocaml}

A função, acima representada, demonstra a definição de um autómato no programa \textit{OCaml}. Cada autómato é composto pelo estado inicial, transições e estados finais. A diferença para a biblioteca \textit{Cytoscape.js} é que, nesta, são compostos só por estados e transições, sendo que as transições dependem dos estados. 

A análise do código explicativo de como gerar graficamente os autómatos será feito de uma forma \textit{down-top}, porque, no \textit{OCaml} funções chamadas por outra devem ser colocadas acima desta.
\begin{ocaml}
let createNode (f, s, t) =
    let test = last1 f in 
    js_run ("makeNode1('"^f^"', '"^string_of_bool (test)^"')");
    let test = last1 t in 
    js_run ("makeNode1('"^t^ "', '"^string_of_bool (test)^"')")
let createEdge (f, s, t) = 
    js_run ("makeEdge('"^f^"', '"^t^"', '"^(String.make 1 s)^"')");
let rec findtransitions (trans) =
    match trans with 
        [] -> false 
        | x::xs -> createNode(x); createEdge (x); findtransitions (xs)
let generateAutomata (gra) =
    automata := gra;
    js_run ("start()");
    ignore (findtransitions (gra.transitions))
\end{ocaml}

\begin{figure}[htbp]
\includegraphics[width=1.0\linewidth]{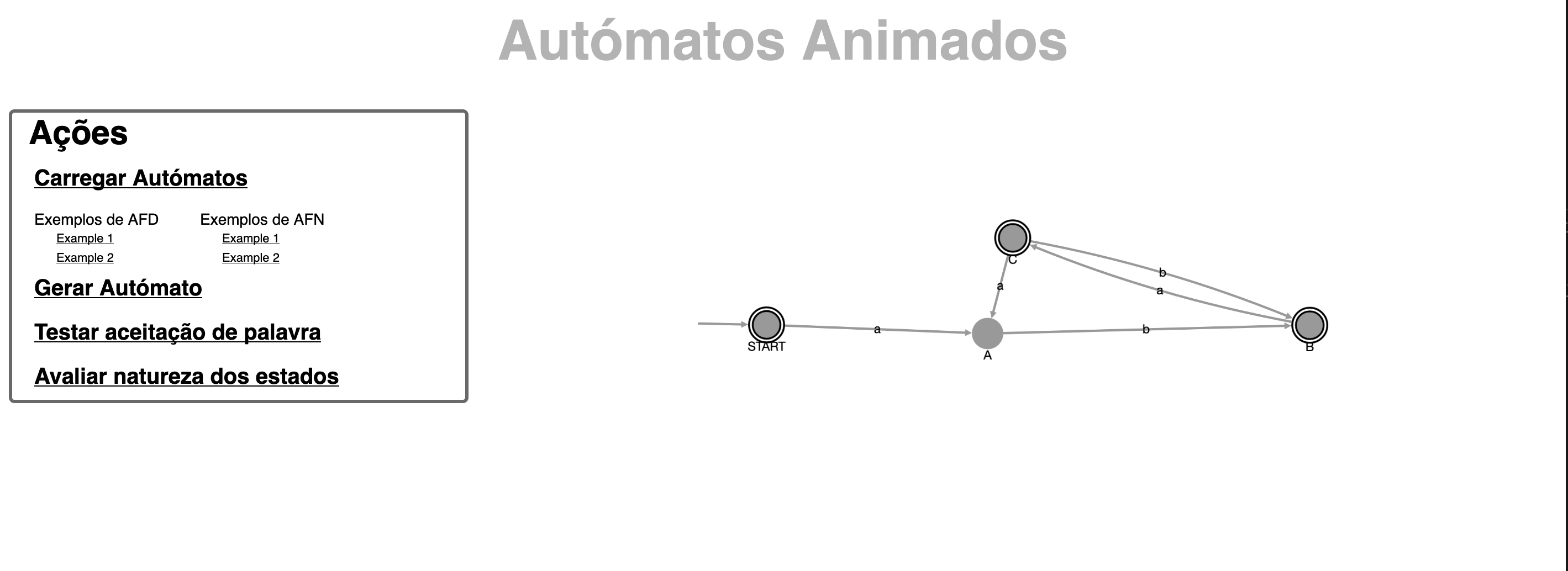}
\caption{Carregar Autómatos - Exemplo1 de AFD}
\label{fig:exemplo1AFD}
\end{figure}

\textit{generateAutomata} é a função chamada para iniciar a criação do autómato. Em primeiro lugar, define a referência \textit{automata} como o autómato a ser gerado. Esta referência serve para que todos as outras funções possam saber qual o autómato representado na página (importante para as funções explicadas nas subsecções seguintes). Em seguida, é chamada a função \textit{start()} do \textit{JavaScript}, que define o tipo de grafo e as suas características principais. Por fim, é chamada a função \textit{findtransitions}, função recursiva que, para cada transição em \textit{Ocaml}, manda criar os estados e a transição correspondentes no \textit{JavaScript}. Terminado este processo de criação do autómato pode, então, visualizar-se um grafo como o da Figura \ref{fig:exemplo1AFD}, que corresponde à representação \textit{OCaml} anteriormente definida.

\subsection{Gerar Autómatos}
\begin{figure}[htbp]
  \centering
  \begin{subfigure}[b]{0.4\textwidth}
    \includegraphics[width=1.0\linewidth]{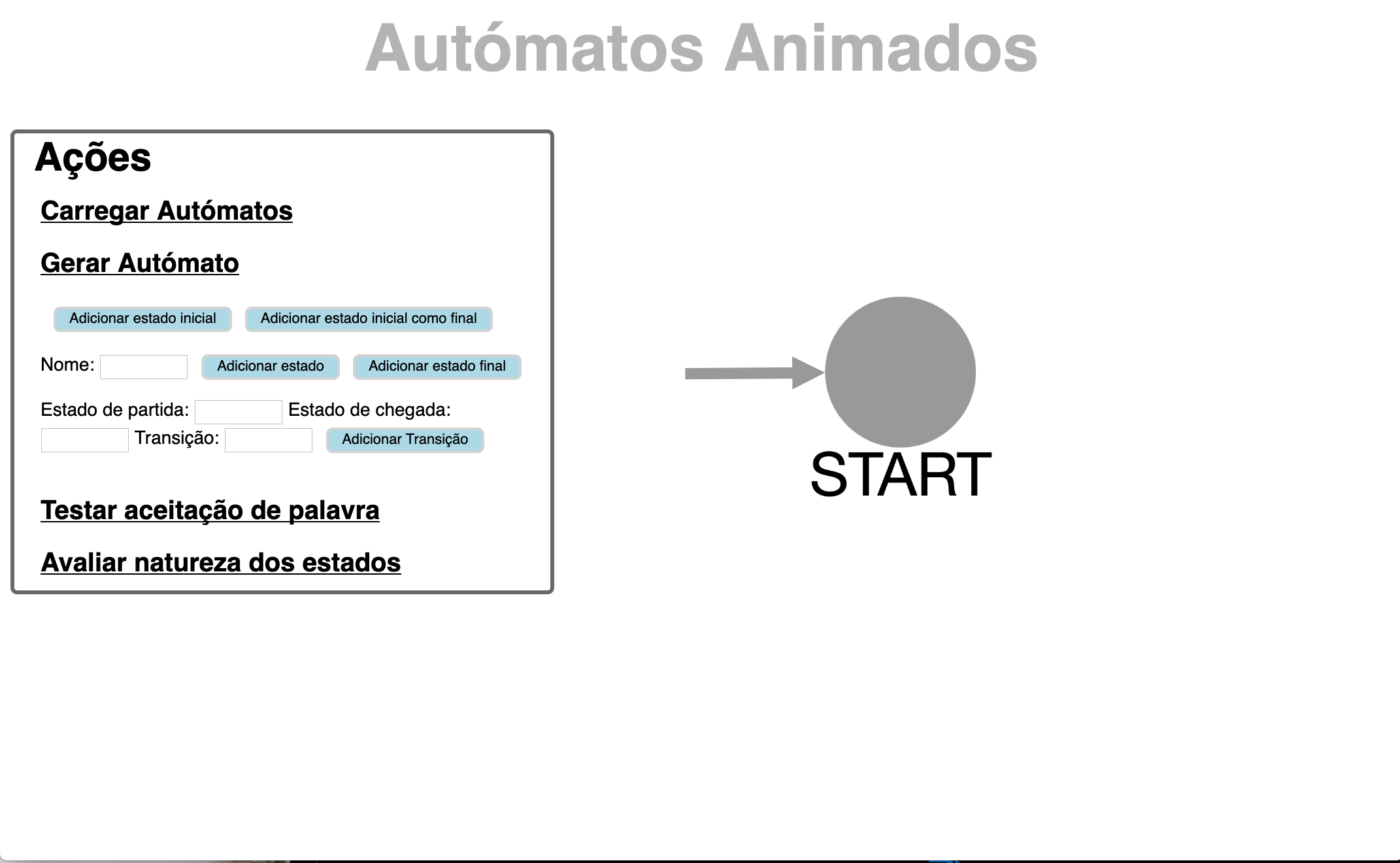}
     \caption{Adicionar estado inicial}
     \label{fig:start}
  \end{subfigure}
  \begin{subfigure}[b]{0.4\textwidth}
    \includegraphics[width=1.0\linewidth]{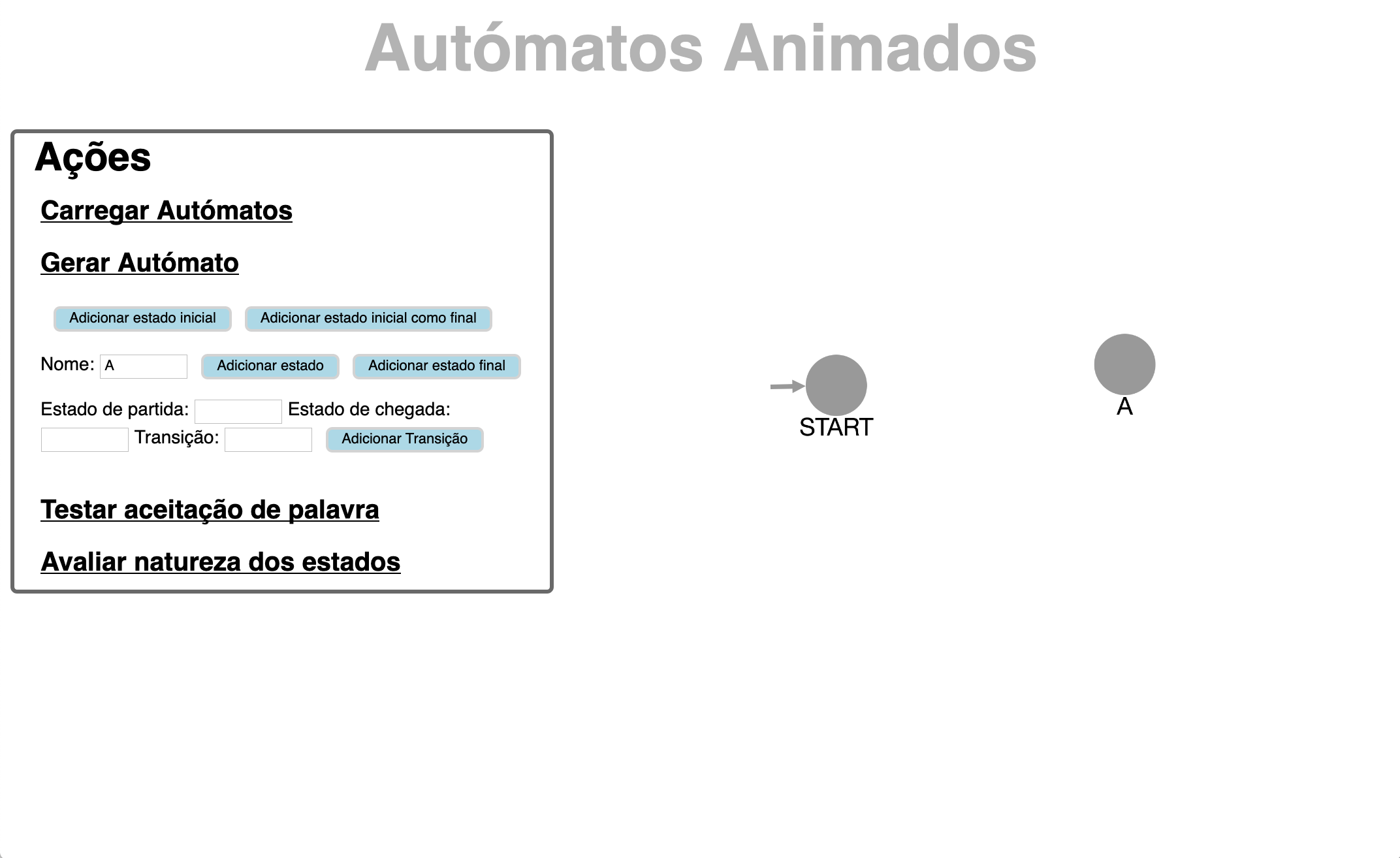}
    \caption{Adicionar A}
    \label{fig:a}
  \end{subfigure}
  \begin{subfigure}[b]{0.4\textwidth}
    \includegraphics[width=1.0\linewidth]{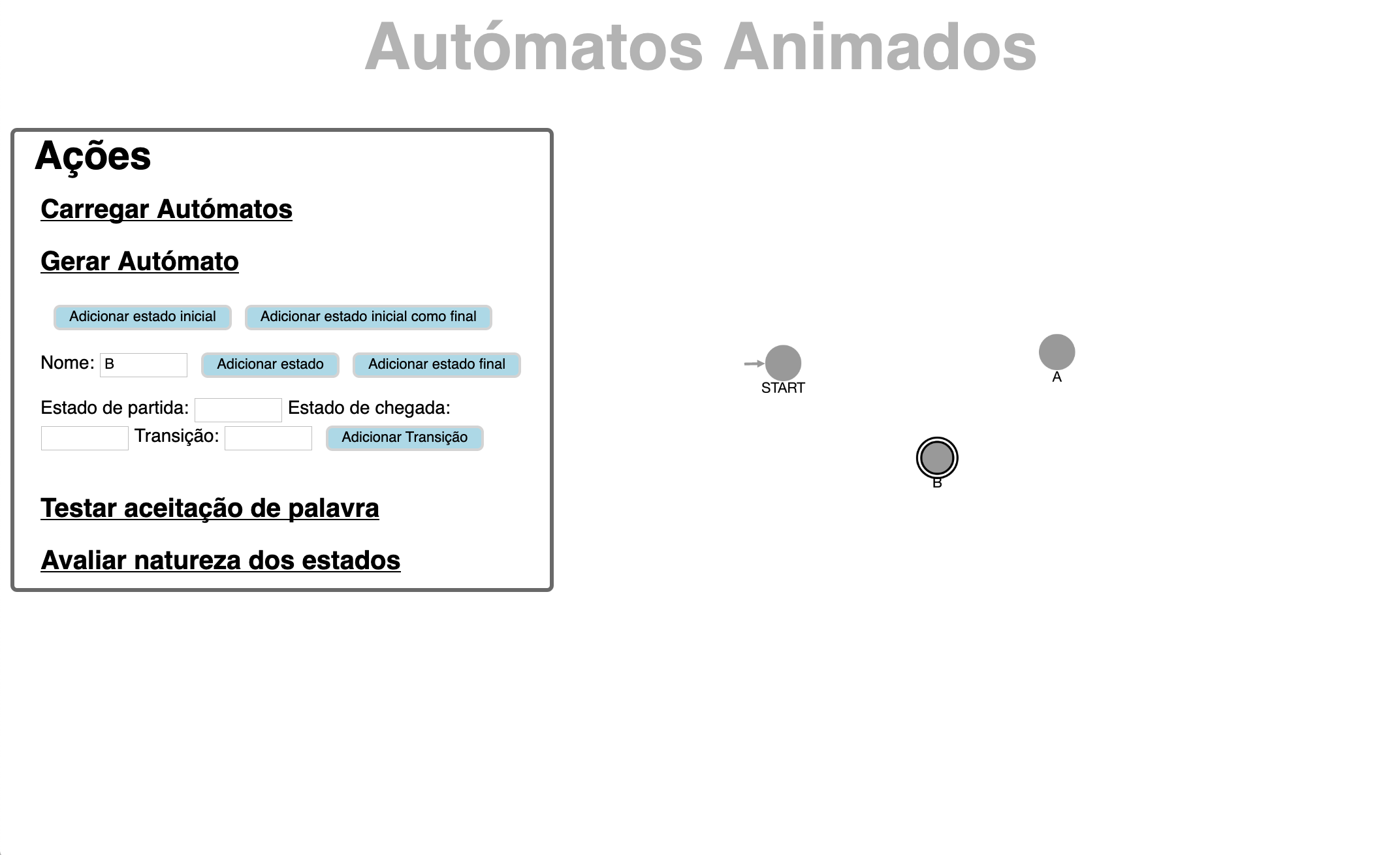}
    \caption{Adicionar B}
    \label{fig:b}
  \end{subfigure}
  \begin{subfigure}[b]{0.4\textwidth}
    \includegraphics[width=1.0\linewidth]{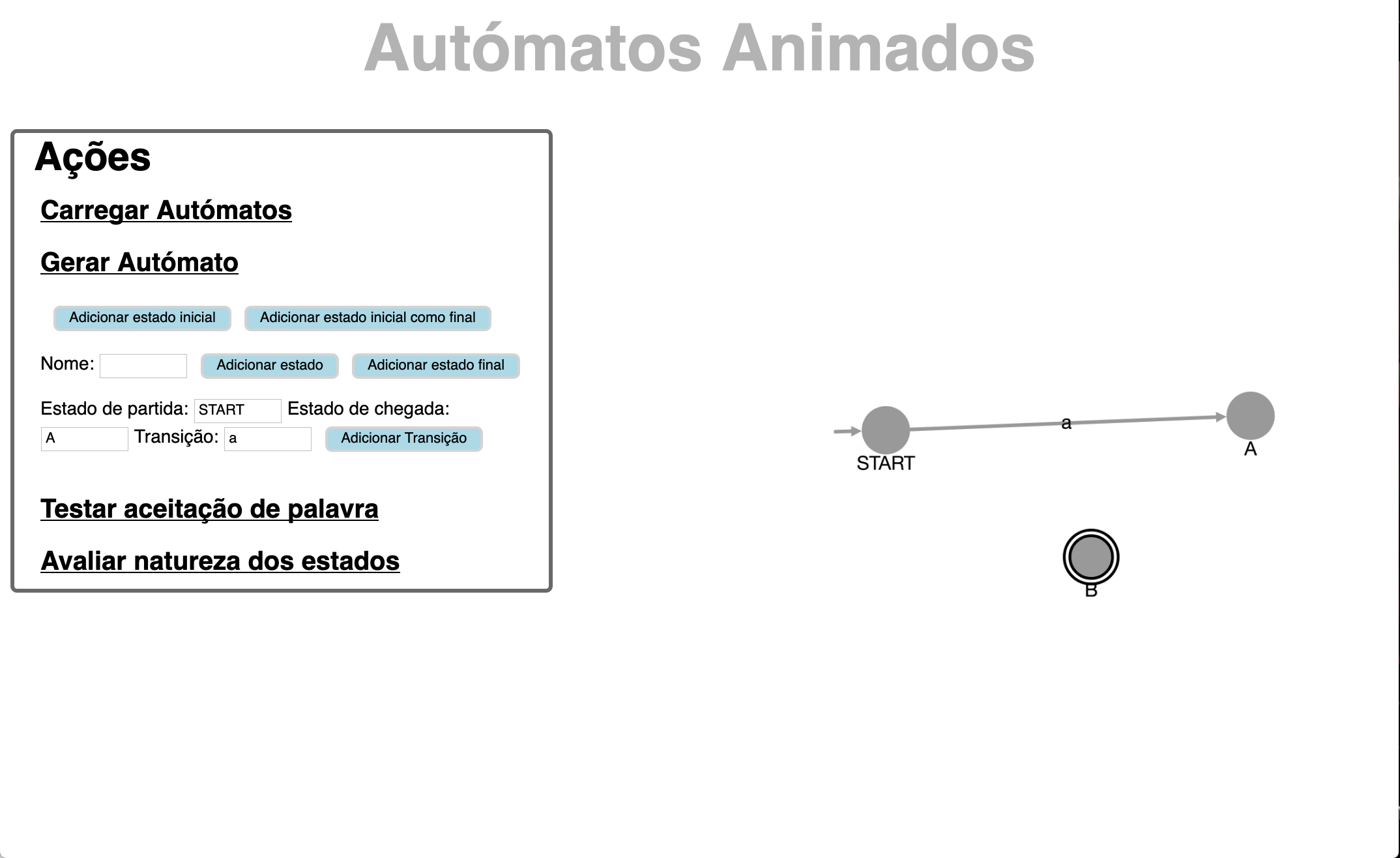}
    \caption{Adicionar Transição}
    \label{fig:transicao}
  \end{subfigure}
  \caption{Geração de um Autómato}
  \label{fig:criacao}
\end{figure}

A opção surge no seguimento da anterior, com o objetivo de permitir ao utilizador criar um autómato do zero (exemplo, Figura \ref{fig:criacao}) ou acrescentar estados e transições aos autómatos carregados. Posto isto, a questão que se coloca é: Como permitir a criação de Autómatos por etapas?

Para  a resolução deste problema foi necessário criar vários botões e caixas de input, para que o utilizador possa identificar os estados e as transições. Para criar um estado é necessário dar input do nome do mesmo e para e para criar uma transição é necessário indicar o estado de partida, o estado de chegada e o símbolo de transição.

Esta fase não foi fácil pois, apesar do framework ter online uma API muito completa esta, por falta de explicações e de exemplos, não é fácil de compreender tornando-se pouco pedagógica.

\begin{figure}[htbp]
  \centering
  \begin{subfigure}[b]{0.75\textwidth}
    \includegraphics[width=1.0\linewidth]{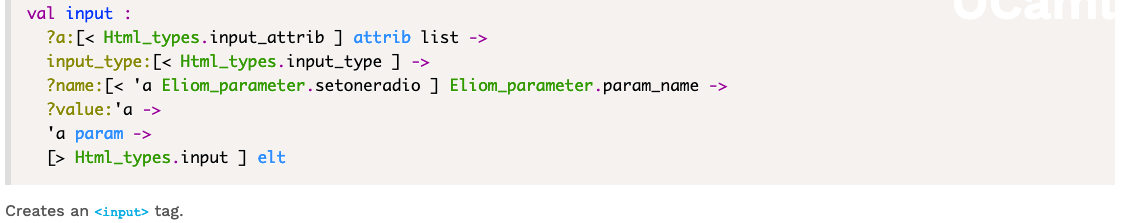}
     \caption{API}
     \label{fig:IputAPI}
  \end{subfigure}
  \begin{subfigure}[b]{0.75\textwidth}
    \includegraphics[width=1.0\linewidth]{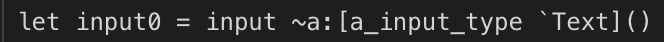}
    \caption{Código caixa de input}
    \label{fig:InputCode}
  \end{subfigure}
  \caption{API e código de uma caixa de input}
  \label{fig:Input}
\end{figure}

A título de exemplo veja-se a Figura \ref{fig:IputAPI}, que mostra o ponto da API, que demonstra como criar uma caixa de \textit{input}. Para se perceber a informação necessária em cada componente é preciso fazer uma nova pesquisa, de forma a compreender como ela é declarada (Figura \ref{fig:InputCode}). Com a falta de exemplos, nem sempre é fácil encontrar a resposta.

Após alguma investigação e experimentação foi possível obter-se o código funcional para a criação de caixas de input e botões como se pode ver no código seguinte:
\begin{ocaml}
let input1 = input ~a:[a_id "box"; a_input_type `Text]() in
let onclick_handler2 = [\%client (fun _ ->
    let i = (Eliom_content.Html.To_dom.of_input ~\%input1) in
    let v = Js_of_ocaml.Js.to_string i##.value in
    js_run ("makeNode1('" ^ v ^ "', '" ^ string_of_bool (false) ^ "')")
  )] in
  let button2 = button ~a:[a_onclick onclick_handler2] [txt "Adicionar estado"]
\end{ocaml}

No código apresentado pode visualizar-se uma característica importante do \textit{framework}, a chamada de código de cliente em código de servidor. As duas funções apresentadas fazem parte do servidor, pois este é que cria os a página web, mas chama código de cliente (representado por [\%cliente]) pois este têm de estar disponível durante todo o funcionamento da página.

Acrescentar estados ou transições ao autómato gerado estava, em parte, resolvido, uma vez que para o  desenhar graficamente era só necessário chamar as funções \textit{JavaScript}, já criadas para o efeito. Como a representação \textit{OCaml} do autómato não pressupõe a representação de estados, o botão para os gerar implica apenas chamar a função \textit{makeNode} do \textit{JavaScript}. Para o autómato ficar atualizado também no \textit{OCaml} criou-se uma função que acrescentasse transições à representação do mesmo.
\begin{ocaml}
let newNode(c1,c2,c3) = {initialState = !automata.initialState;
                        transitions = !automata.transitions@[(c1,c2,c3)];
                        acceptStates = !automata.acceptStates}
\end{ocaml}

Para gerar um autómato novo, criado de raiz, foi necessário criar um botão que inicializasse a biblioteca e colocasse o estado de partida na página. Este botão foi mais tarde duplicado para permitir que o estado de partida pudesse ser também final.

Por fim, para a geração de estados foi também criado um novo botão para identificar o estado como final. Este botão pressupôs a criação de uma função que o acrescentasse à representação \textit{OCaml} como estado final.
\begin{ocaml}
let newNodeFinal final = {initialState = !automata.initialState;
                          transitions = !automata.transitions;
                          acceptStates = !automata.acceptStates@[final]}
\end{ocaml}

\subsection{Testar palavras}

\begin{figure}[htbp]
  \centering
  \begin{subfigure}[b]{0.4\textwidth}
    \includegraphics[width=1.0\linewidth]{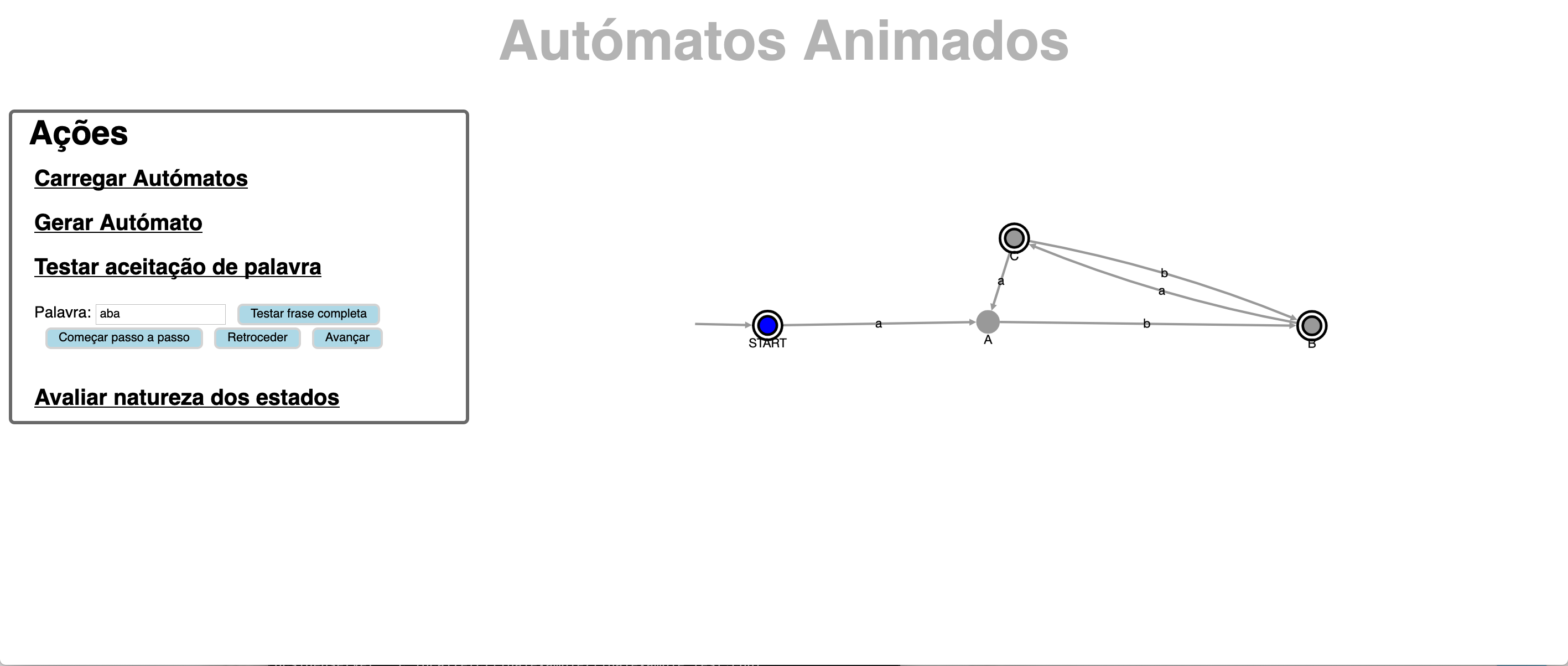}
     \caption{Estado Inicial}
     \label{fig:inicial1}
  \end{subfigure}
  \begin{subfigure}[b]{0.4\textwidth}
    \includegraphics[width=1.0\linewidth]{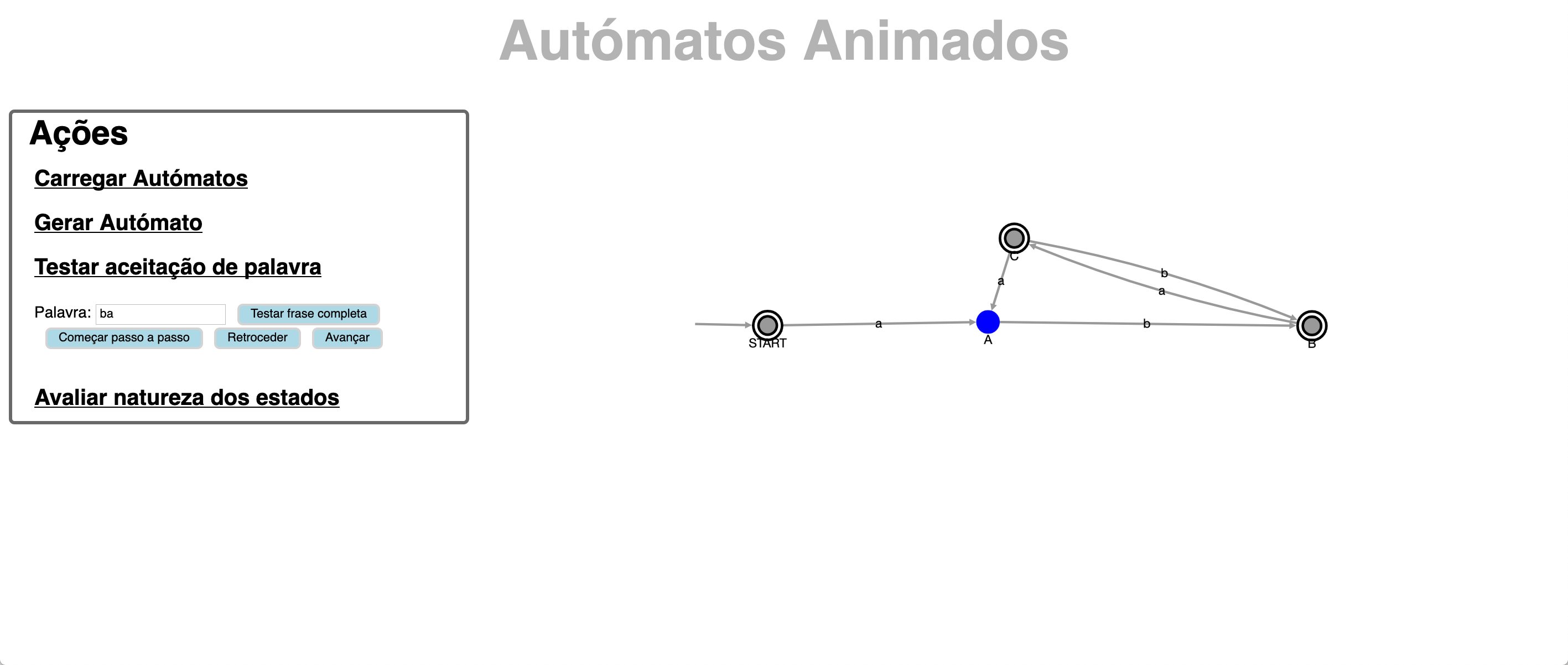}
    \caption{Segundo estado}
    \label{fig:segundo1}
  \end{subfigure}
  \begin{subfigure}[b]{0.4\textwidth}
    \includegraphics[width=1.0\linewidth]{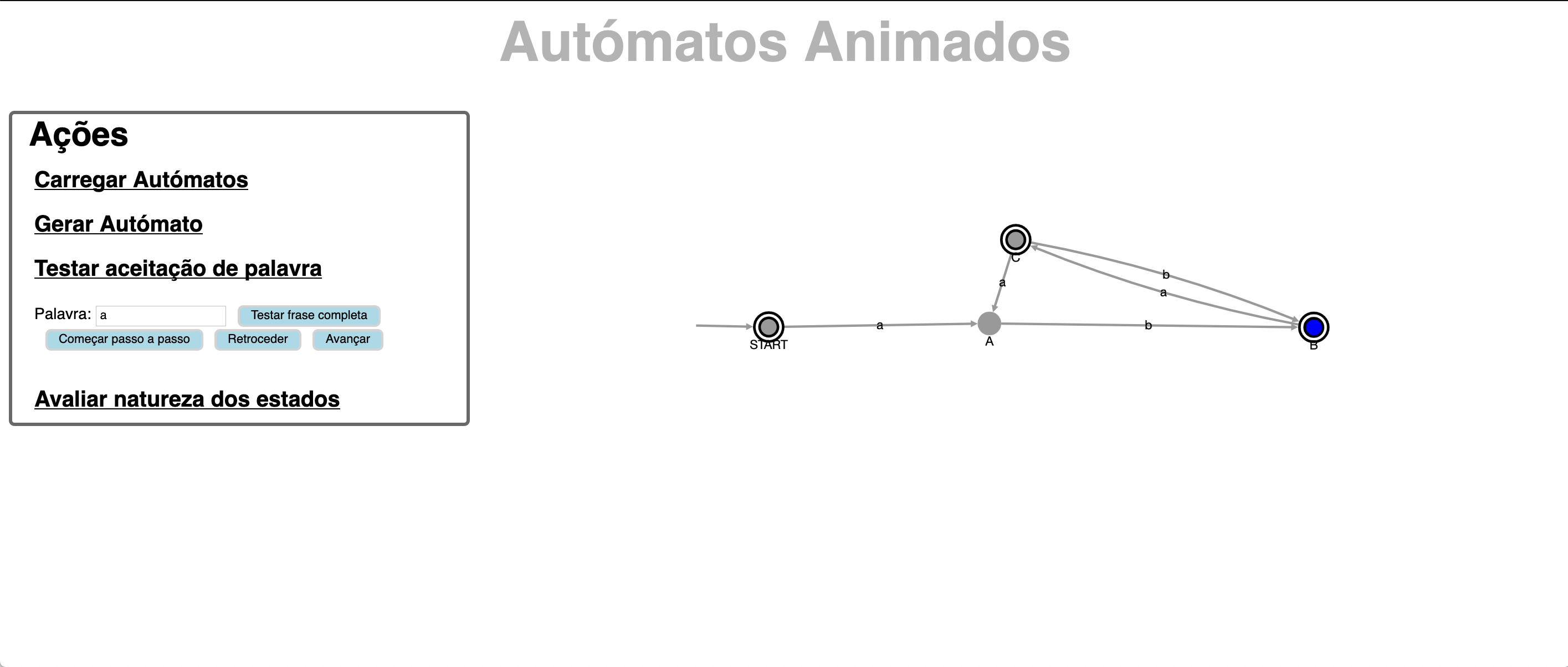}
    \caption{Terceiro Estado}
    \label{fig:terceiro1}
  \end{subfigure}
  \begin{subfigure}[b]{0.4\textwidth}
    \includegraphics[width=1.0\linewidth]{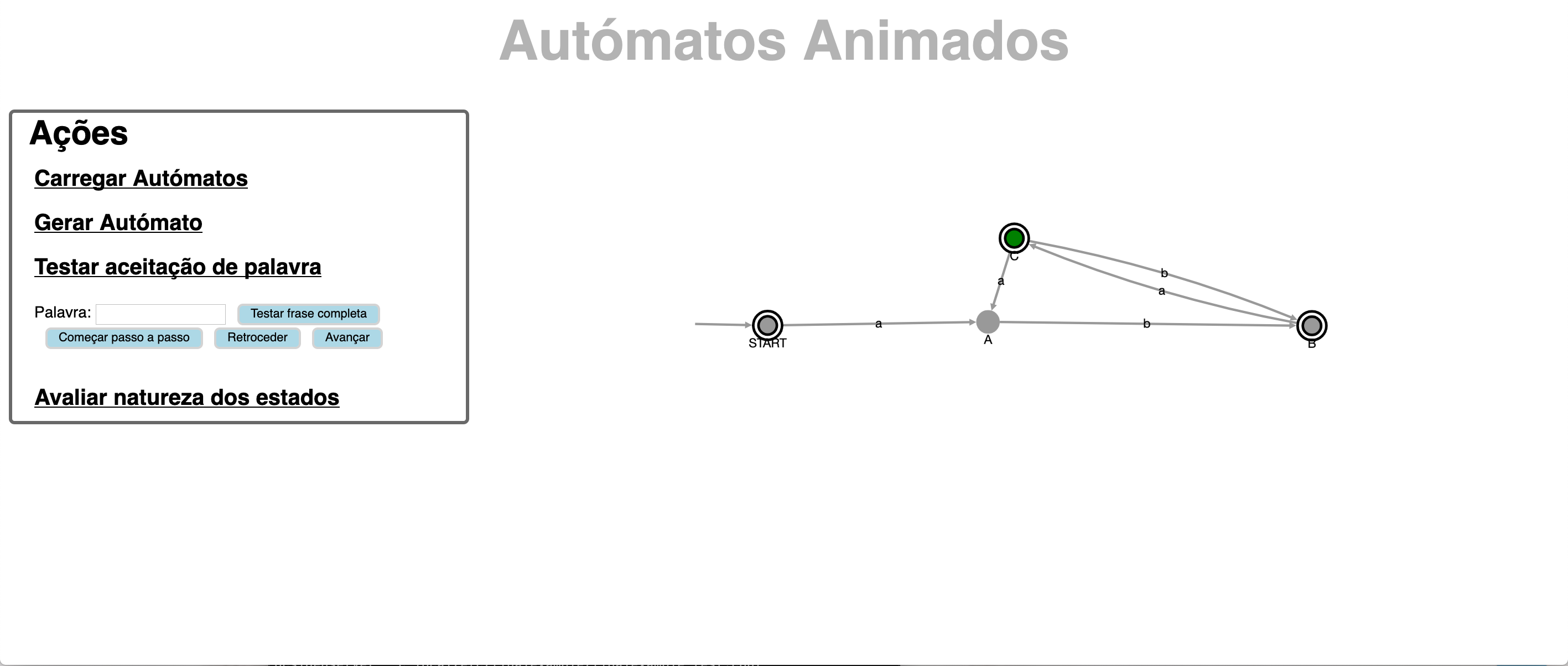}
    \caption{Estado Aceite}
    \label{fig:aceite}
  \end{subfigure}
  \caption{Verificação da aceitação da palavra aba - palavra aceite}
  \label{fig:aceitacao1}
\end{figure}

\begin{figure}[htbp]
  \centering
  \subcaptionbox{Quarto estado \label{fig:quarto2}}%
    {\includegraphics[width=0.5\linewidth]{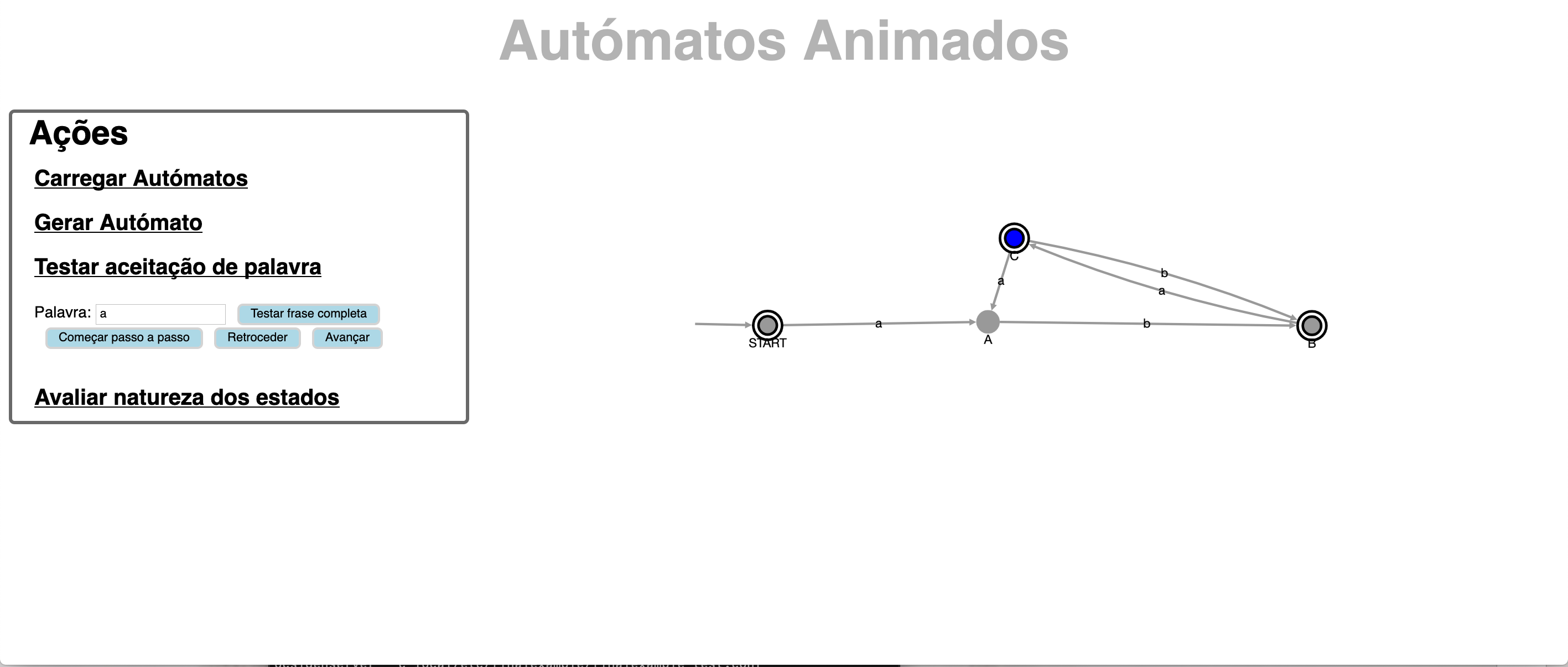}}%
  \subcaptionbox{Estado final: Erro \label{fig:erro2}}%
    {\includegraphics[width=0.5\linewidth]{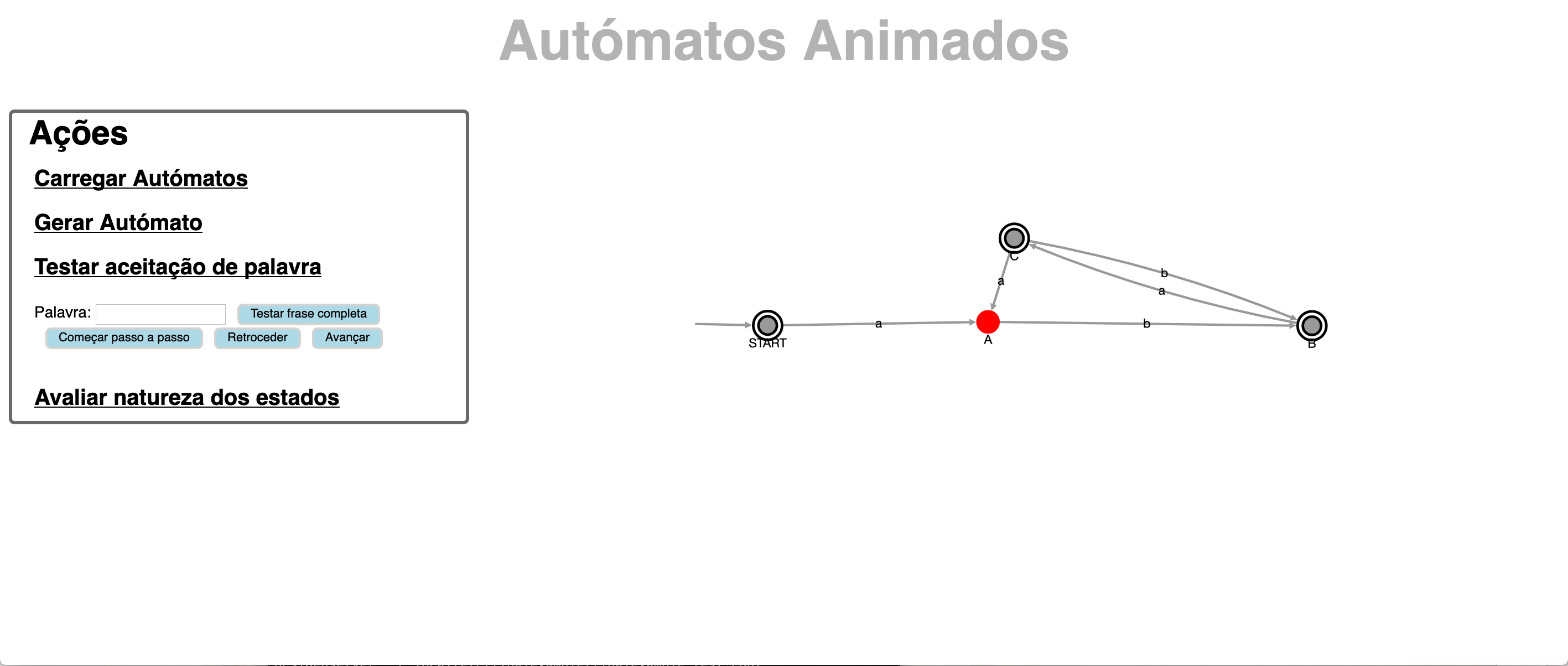}}%
  \caption{Verificação da aceitação da palavra abaa - palavra não aceite (começo igual a Figuras \ref{fig:inicial1}, \ref{fig:segundo1}, \ref{fig:terceiro1})}
  \label{fig:aceitacao2}
\end{figure}

A opção testar palavras foi a de desenvolvimento mais desafiante e, por enquanto, funciona só para AFDs e alguns AFNs. Pode visualizar-se um exemplo da aceitação da palavra nas Figuras \ref{fig:aceitacao1} e \ref{fig:aceitacao2}. A ideia base é simples: o utilizador faz input de uma frase que quer testar e, escolhendo uma das duas opções, terá oportunidade de ver os estados a mudar de cor: azul, se é um estado de passagem, vermelho, se chegou ao fim da palavra mas não ficou num estado final (palavra não aceite) ou verde, se a palavra chegou ao fim e o estado é final (palavra aceite). Partindo do exemplo falado na Subsecção anterior, foi possível criar uma caixa de input e um botão que lê a informação da caixa. Foi, em seguida, necessário responder a três perguntas: Como editar o autómato? Como animar o autómato? E como fazer a animação passo a passo? \smallskip

\textbf{Como editar o autómato?} Como referido, a edição do autómato é feita no código \textit{Javascript}, mas a lógica da aplicação está no \textit{OCaml}. Assim, o importante era perceber como editar o estilo do grafo (problema resolvido com o estudo da biblioteca) e quais os dados que devem ser passados aquando da chamada da função \textit{JavaScript}. A informação necessária é o estado para mudar a cor, o ponto da palavra em que nos encontramos e se o estado é final. \smallskip

\textbf{Como animar o autómato?} Para que o utilizador tenha a possibilidade de visualizar a passagem dos estados é necessário animar o autómato gerado. Criou-se, assim, uma função que retorna um estado ao qual se chega com o símbolo dado, partindo do estado atual. Quando um novo estado é encontrado, este é colorido, chamando a função \textit{JavaScript} para o efeito. Após esta construção percebeu-se que apenas o estado final era colorido. 

Para dar tempo à página de modificar o autómato recorreu-se à biblioteca \textit{OCaml Unix}, que contém uma função \textit{sleep}. Problema: a função está escrita na parte de cliente e a biblioteca só funciona no servidor. Depois de alguma pesquisa percebeu-se que a solução seria utilizar a biblioteca de \textit{threads Lwt} e uma função de \textit{delay}. Obteve-se o seguinte:
\begin{ocaml}
let rec delay  n = if n = 0 then Lwt.return () else Lwt.bind (Lwt.pause ()) (fun () -> delay (n-1))
let rec acceptX s w fa =
  let last = last1 s in 
  js_run ("changeColor1('"^s^"', '"^(string_of_int (List.length w))^"', '"^ string_of_bool (last)^"')");
  match w with
	[] -> Lwt.return (List.mem s fa.acceptStates)
	| x::xs -> match transitionsFor (s,x) fa with
           [] -> Lwt.return (js_run ("changeColor1('"^s^"', '"^(string_of_int (0))^"', '"^string_of_bool (false)^"')"); false)
           | (_,_,s)::_ -> Lwt.bind (delay 50)
                            (fun () -> Lwt.bind (Lwt.return (let last3 = last1 s in js_run ("changeColor1('"^s^"', '"^(string_of_int (List.length xs))^"', '"^string_of_bool (last3)^"')"))) (fun () -> acceptX s xs fa))
\end{ocaml}

O que acontece é que cada vez que se encontra uma transição, é executado um \textit{delay} onde é usada a função \textit{Lwt} de pausa que, trata os eventos pendentes, incluindo os de atualização do ecrã. \smallskip

\textbf{E como fazer a animação passo a passo?} Para se desenvolver a opção passo a passo foi necessário criar três novas variáveis imperativas: \textit{position, step} e \textit{sentence}. \textit{Position} indica em que ponto da palavra nos encontramos: quando se anda para a frente esta posição é incrementada, verificando-se o oposto quando se anda para trás. \textit{Step} indica o último estado por onde se passou e \textit{sentence} representa a lista de símbolos presentes na palavra.

Existe um botão para começar a iteração, que pinta o estado inicial, e inicializa as variáveis mencionadas.

Partindo da função recursiva de animação do autómato, foi possível, agora sem a recursividade e sem a utilização da biblioteca \textit{Lwt}, criar uma função que retorna o estado seguinte a um dado estado (\textit{step}) consoante o símbolo em dada posição (\textit{position}) da palavra a ser testada (\textit{sentence}). Conseguimos, assim, a função de avançar.

Para realizar a função de retroceder, foi necessário fazer uma modificação à função de inicialização. Esta última passa a chamar outra que cria uma lista ordenada de todos os estados que  vão ser coloridos. Assim, quando se clica no botão "retroceder", a posição (variável \textit{position}) é decrementada e o estado correspondente à mesma na lista é colorido.
\begin{ocaml}
let acceptStepBack w =
  if (!position > 0) then 
    (position := !position - 1; 
    let st = get_nth !listStates !position in 
    let pos = List.length w - !position in 
    let isFinal = last1 st in 
    js_run ("changeColor1('"^st^"', '"^string_of_int (pos)^"', '"^ string_of_bool (isFinal)^"')"); step := setStep st)
\end{ocaml} 

É importante referir que aquando do clique nos botões "avançar" ou "retroceder" é chamada uma função que atualiza a frase na caixa de input para a situação em que o utilizador se encontra.

\subsection{Verificar Natureza}

\begin{figure}[htbp]
\includegraphics[width=1.0\linewidth]{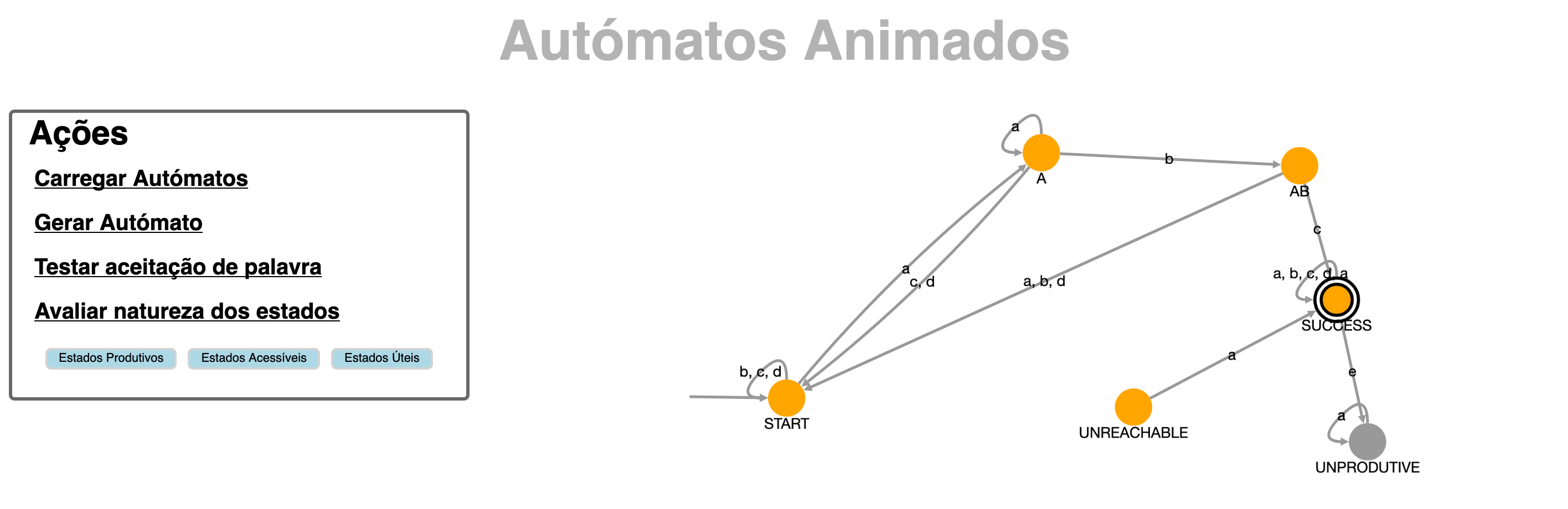}
\caption{Indicação dos estados produtivos}
\label{fig:produtivos}
\end{figure}

\begin{figure}[htbp]
\includegraphics[width=1.0\linewidth]{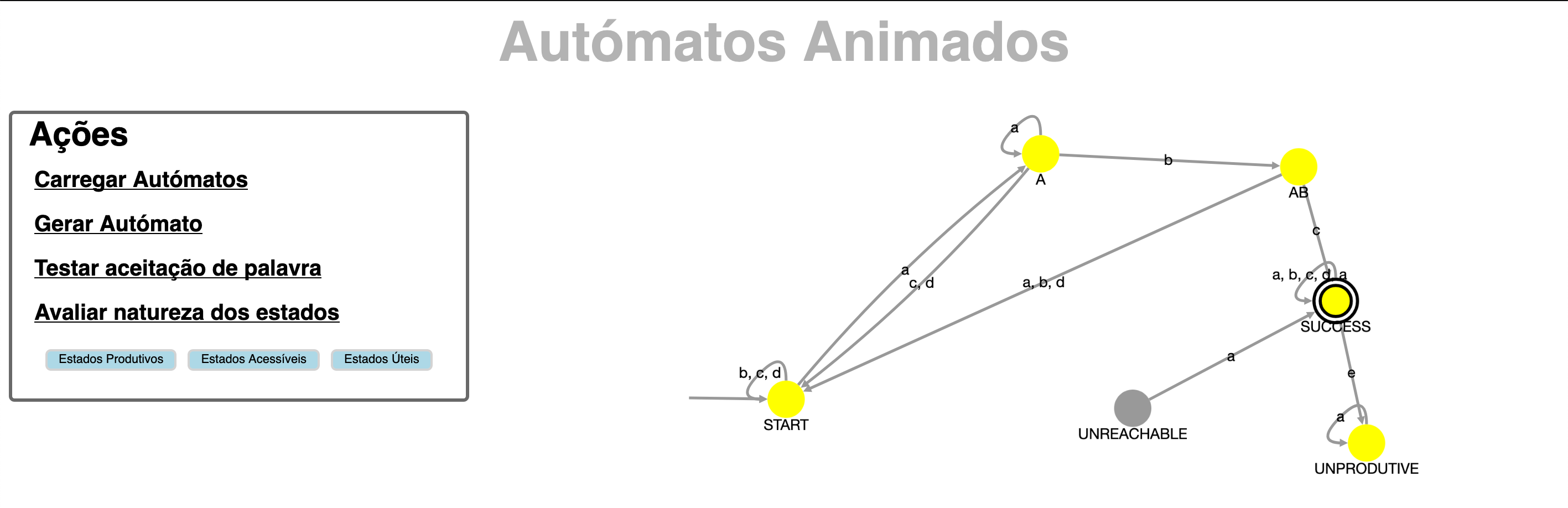}
\caption{Indicação dos estados acessíveis}
\label{fig:acessiveis}
\end{figure}

\begin{figure}[htbp]
\includegraphics[width=1.0\linewidth]{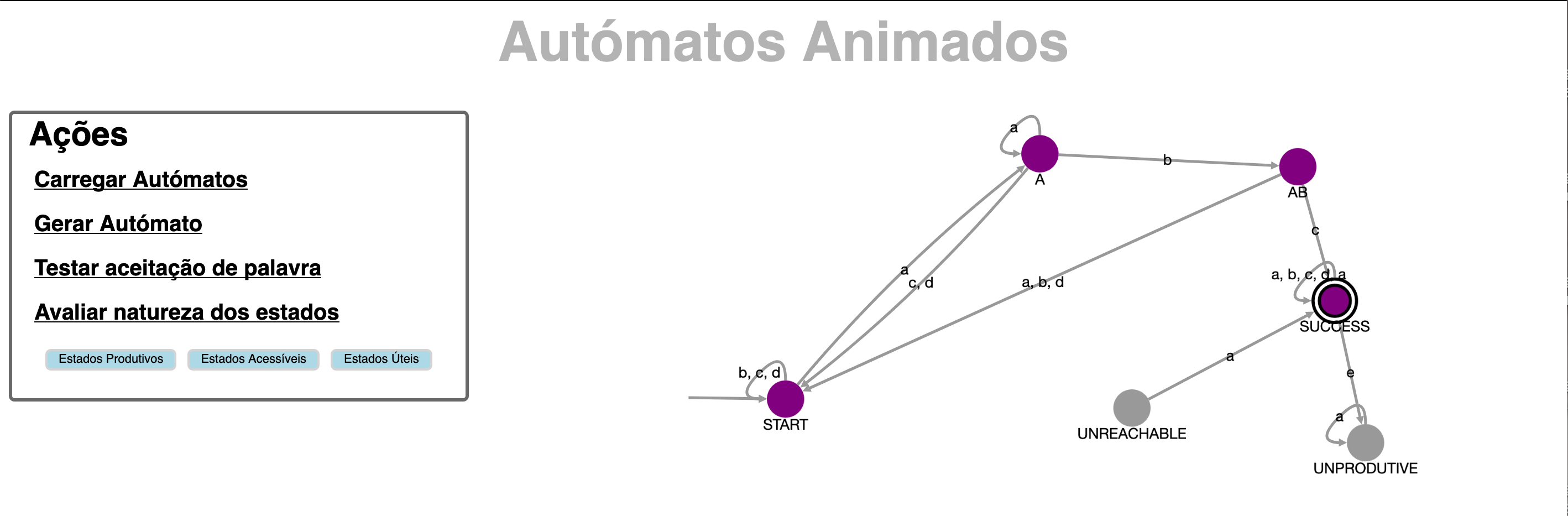}
\caption{Indicação dos estados úteis}
\label{fig:uteis}
\end{figure}

Esta opção pretende que o utilizador perceba se os estados são produtivos, acessíveis ou úteis. Para cada uma das opções existe um botão que assinala, em simultâneo, todos os estados que correspondem a essa característica. A questão que surgiu foi: Como editar graficamente um estado sem alterar os outros? Anteriormente era feito um \textit{reset} do estilo do autómato a cada afectação, neste caso foi necessária uma pesquisa na API da biblioteca para perceber como fazer a sua atualização. Em JavaScript, para cada uma das opções, foi criada uma função que colorisse um dado nó sem modificar os outros. Em \textit{OCaml} criaram-se diferentes funções, uma para encontrar os estado produtivos, outra para encontrar os estados acessíveis e por fim uma que intersectasse os dois resultados (estados úteis). Ao selecionar uma das opções é chamada a função \textit{OCaml} correspondente, sendo que cada estado obtido é colorido através da função \textit{JavaScript} respetiva. Nas Figuras \ref{fig:produtivos}, \ref{fig:acessiveis} e \ref{fig:uteis} é podem ver-se exemplos de cada uma das opções.
\ifshortversion
\section{Conclusões e Trabalhos Futuros}
\fi
Utilizando o \emph{framework} \textit{Ocsigen}, está-se a desenvolver uma aplicação web interactiva para apoiar o ensino de Linguagens Formais e Autómatos. A vantagem do  \textit{Ocsigen} é permitir desenvolver de forma integrada tando a parte do servidor como do cliente, numa linguagem concisa e segura como o \textit{OCaml}.



Um aspecto central da plataforma é ser extensível;
pretende-se incluir
mais funcionalidades e outras classes de linguagens, bem como integrar com suporte à avaliação, nomeadamente permitindo a submissão e classificação de exercícios.





\bibliographystyle{splncs04}
\bibliography{bibliography.bib}

\end{document}